\documentclass[letterpaper]{article} 
\usepackage[]{aaai2026}  
\usepackage{times}  
\usepackage{helvet}  
\usepackage{courier}  
\usepackage[hyphens]{url}  
\usepackage{graphicx} 
\usepackage{natbib}  
\usepackage{caption} 
\usepackage{algorithm}
\usepackage{algorithmic}

\usepackage{graphicx}
\usepackage{amsmath}
\usepackage{amssymb}
\usepackage{booktabs}
\usepackage{natbib}
\usepackage{algorithm}
\usepackage{algorithmic}
\usepackage{multirow}
\usepackage{pifont}     
\usepackage[table]{xcolor} 
\usepackage{subcaption}
\usepackage{graphicx, booktabs, multirow} 

\usepackage{tabularx}
\usepackage{booktabs}
\usepackage{siunitx}

\usepackage{newfloat}
\usepackage{listings}
\DeclareCaptionStyle{ruled}{labelfont=normalfont,labelsep=colon,strut=off} 
\floatstyle{ruled}
\newfloat{listing}{tb}{lst}{}
\floatname{listing}{Listing}
\DeclareRobustCommand{\logo}{%
  \begingroup\normalfont
  \raisebox{-0.25em}{%
  \hspace{-0.5em}
  \includegraphics[height=1.5em]{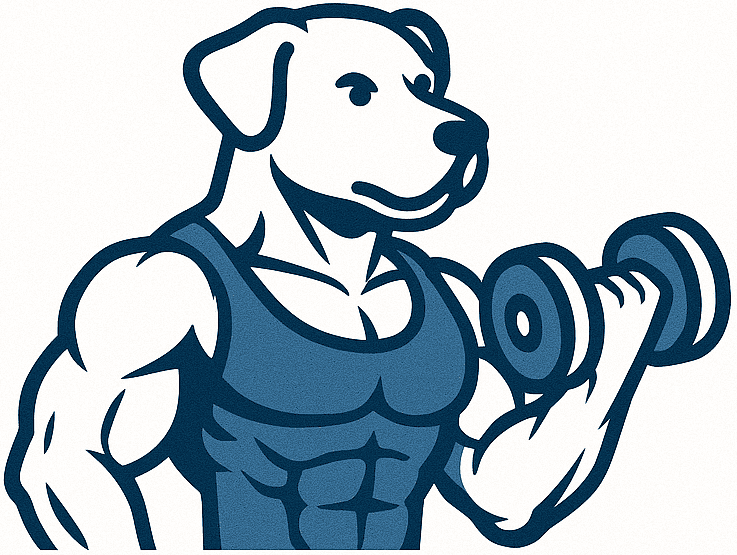}%
  }%
  \kern 0.2em
  \endgroup
}

\title{\logo DogFit: \underline{Do}main-\underline{g}uided \underline{Fi}ne-\underline{t}uning for Efficient Transfer Learning of Diffusion Models}

\author {
    Yara Bahram,
    Mohammadhadi Shateri,
    Eric Granger
}
\affiliations {
    LIVIA, ILLS, ETS Montreal, Canada\\
    yara.mohammadi-bahram@livia.etsmtl.ca, \{mohammadhadi.shateri, eric.granger\}@etsmtl.ca
}

\begin{document}

\maketitle\begin{abstract}
Transfer learning of diffusion models to new domains with limited data is challenging, as naively fine-tuning the model often results in poor generalization. Test-time guidance methods help mitigate this by offering controllable improvements in image fidelity through a trade-off with sample diversity. However, this benefit comes at a high computational cost, typically requiring dual forward passes during sampling.  We propose the \underline{Do}main-\underline{g}uided \underline{Fi}ne-\underline{t}uning (\texttt{DogFit}) method, an effective guidance mechanism for diffusion transfer learning that maintains controllability without incurring additional computational overhead. \texttt{DogFit} injects a domain-aware guidance offset into the training loss, effectively internalizing the guided behavior during the fine-tuning process. The domain-aware design is motivated by our observation that during fine-tuning, the unconditional source model offers a stronger marginal estimate than the target model. To support efficient controllable fidelity–diversity trade-offs at inference, we encode the guidance strength value as an additional model input through a lightweight conditioning mechanism. We further investigate the optimal placement and timing of the guidance offset during training and propose two simple scheduling strategies, i.e., \textit{late-start} and \textit{cut-off}, which improve generation quality and training stability. Experiments on DiT and SiT backbones across six diverse target domains show that \texttt{DogFit} can outperform state-of-the-art guidance methods in transfer learning in terms of FID and $\text{FD}_{\text{DINOV2}}$ while requiring up to $\sim\times2$ fewer sampling TFLOPS. \\Code is available at https://github.com/yaramohamadi/DogFit.
\end{abstract}



\section{Introduction}
\label{sec:intro}

Denoising diffusion probabilistic models (DDPMs)~\citep{sohl2015deep, ho2020denoising} have emerged as powerful generative models, achieving state-of-the-art (SOTA) results in image synthesis~\citep{dhariwal2021diffusion, rombach2022high}, video generation~\citep{gupta2024photorealistic, ho2022video}, and image editing~\citep{kawar2023imagic}. 
However, producing high-quality and diverse outputs still requires substantial computational and data resources, particularly with large diffusion backbones and image sizes. In practical applications such as personalized generation with limited data, full model training from scratch is infeasible, making transfer learning a critical tool for target adaptation. A common approach in this setting is to fine-tune a pre-trained diffusion model on the target data. While straightforward, this strategy frequently suffers from overfitting and poor generalization due to the limited size and diversity of the target domain.


\begin{figure}[t]
    \centering
    \includegraphics[width=\linewidth]{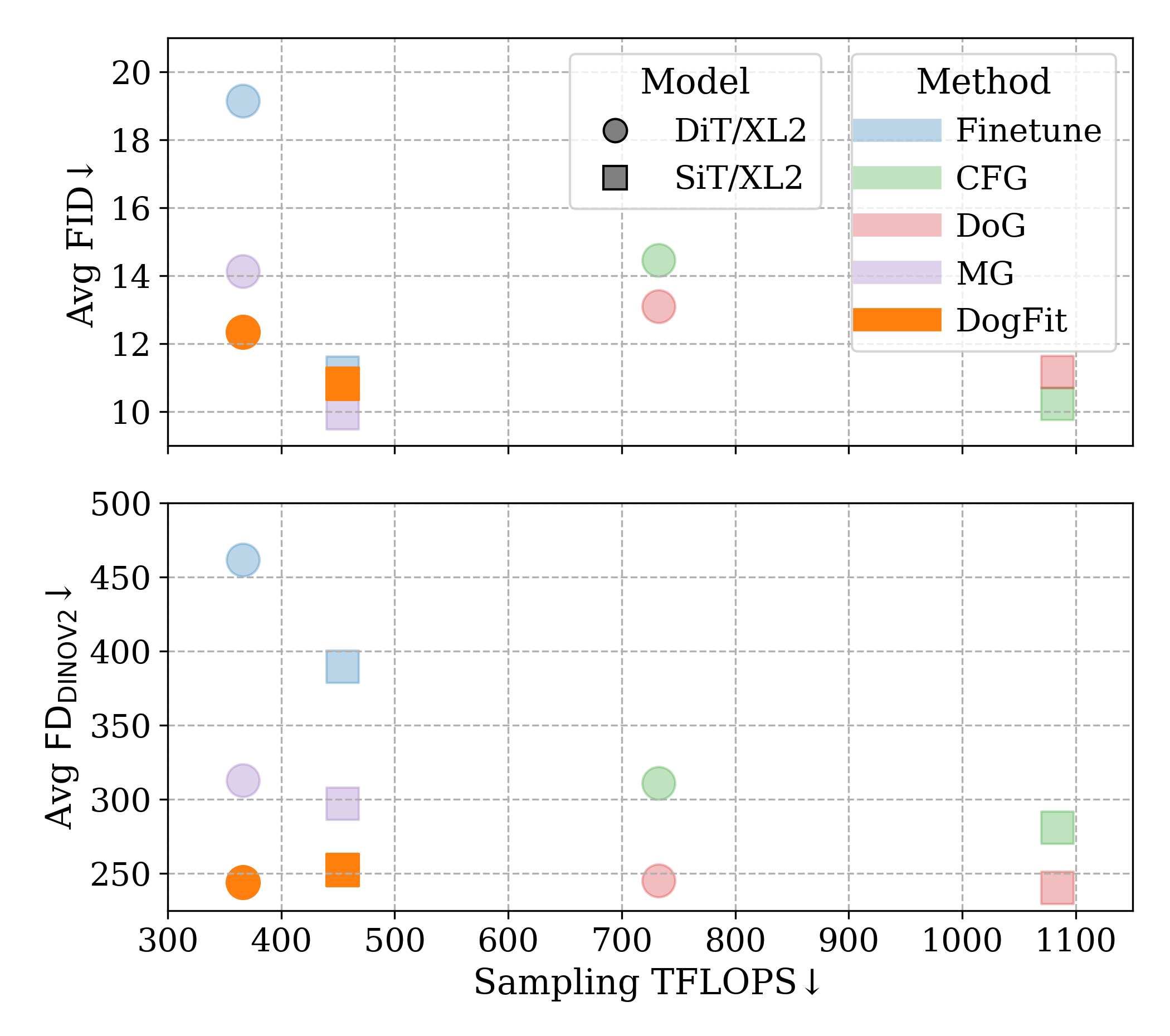}
    \caption{\texttt{DogFit} achieves competitive average FID and $\mathrm{FD}_{\text{DINOV2}}$ when compared to SOTA guidance mechanisms for diffusion transfer learning, all without increasing the computational complexity (sampling TFLOPS). Averages are computed over six target datasets.}
    \label{fig:fid_cub}
\end{figure}
\underline

Classifier-free guidance (CFG) is a standard technique for improving generation quality at inference time by modulating the conditional signal~\citep{ho2022classifier, karras2024guiding}. This technique is often used in its original form even after target adaptation to steer the model towards better generalization. However, CFG assumes access to well-trained conditional and unconditional models, which is difficult to ensure in transfer scenarios where the target domain is small or lacks labels \cite{ho2022classifier, zhong2025domain}. 
Recent methods like domain guidance (DoG) address these issues by utilizing the unconditional source model as a stronger marginal noise estimator, offering improved target domain alignment and generalization ~\citep{zhong2025domain, phunyaphibarn2025unconditional}.
Nevertheless, both CFG and DoG introduce significant computational overhead at sampling time due to their reliance on dual forward passes~\citep{ho2022classifier, zhong2025domain}. On the other hand, model guidance (MG) mitigates CFG's runtime cost in in-domain settings by injecting the guidance signal directly into the diffusion training objective~\citep{chen2025visual, tang2025diffusion}. However, MG inherits the limitations of CFG in transfer learning due to the underfitting of the unconditional noise estimator~\citep{zhong2025domain}. Further, MG hard-codes the guidance strength value during training, restricting control at inference. These limitations raise a key question: \textit{can we design effective guidance mechanisms for diffusion transfer learning without incurring computational overhead and still maintaining controllability over the guidance strength?} 

This paper proposes domain-guided fine-tuning (\texttt{DogFit}), a method that injects a domain-aware guidance offset into the training loss during the fine-tuning process. This allows the diffusion model to directly learn the guided direction, removing the need for additional test-time processes. The domain-aware design is motivated by our observation that during fine-tuning, the source model offers stronger marginal noise estimates than the evolving target model. This design encourages the model to step toward the target domain manifold and avoid out-of-distribution generation. 
To support efficient control over the guidance strength at inference, \texttt{DogFit} conditions the model on the guidance value and exposes it to a range of such values during training. This allows the model to modulate its generation behavior accordingly, enabling fidelity–diversity trade-offs at inference with negligible sampling overhead. 
We further investigate the optimal placement and timing of the guidance offset during training. Two cost-effective guidance scheduling mechanisms are proposed for \texttt{DogFit}: (1) a \textit{late-start} strategy that delays guidance until the model has learned sufficiently stable target representations; and (2) a \textit{cut-off} scheme that restricts guidance to the later denoising steps, where fine-grained domain-specific details are more prevalent. Results over six datasets with different distribution shifts and supervision levels, using two SOTA diffusion backbones, DiT \citep{peebles2023scalable} and SiT \citep{ma2024sit}, show that \texttt{DogFit} can outperform SOTA guidance methods in terms of FID and $\text{FD}_{\text{DINOV2}}$ while reducing their sampling TFLOPS by up to 2× (Fig.\ref{fig:fid_cub}). Results establish \texttt{DogFit} as a practical and efficient guidance method for diffusion transfer learning.



\vspace{0.5em}
\noindent\textbf{Contributions. } \textbf{(i)} We propose \texttt{DogFit}, a training-time guidance mechanism for diffusion transfer learning that enables controllable improvement over target domain generation fidelity without requiring additional processes or double forward passes at test-time. \textbf{(ii)} We show that, during fine-tuning, the source domain model offers stronger marginal estimates than the target model, making it better suited for generating guidance signals. We empirically analyze how the timing and placement of this guidance impact training and propose two scheduling strategies that enhance training stability and generation quality. \textbf{(iii)} Extensive experiments are conducted across diverse target datasets on DiT and SiT diffusion backbones, showing that \texttt{DogFit} can outperform SOTA guidance methods in transfer learning.







\section{Related Work}

\noindent\textbf{Efficient Diffusion Models.}
The iterative denoising process in diffusion models imposes high sampling-time costs~\citep{shen2025efficient}. To accelerate generation, prior work reduces the number of sampling steps via improved solvers~\citep{song2020denoising, lu2022dpm}, optimized noise schedules~\citep{zheng2023improved}, or distillation-based methods that train few(one)-step student models to mimic large multi-step teachers~\citep{salimans2022progressive, yin2024one}. Recent efforts ~\citep{jensen2025efficient, hsiao2024plug, zhou2025dice} distill CFG to eliminate test-time dual passes. Our method, similar to CFG distillation, aims to internalize guidance, but does so without the architectural modifications or post-hoc training stages required by distillation methods. Instead, \texttt{DogFit} simply integrates guidance directly into the fine-tuning objective.

\begin{figure*}[t]
    \centering
    \begin{subfigure}[t]{0.32\textwidth}
        \centering
        \includegraphics[height=5.7cm]{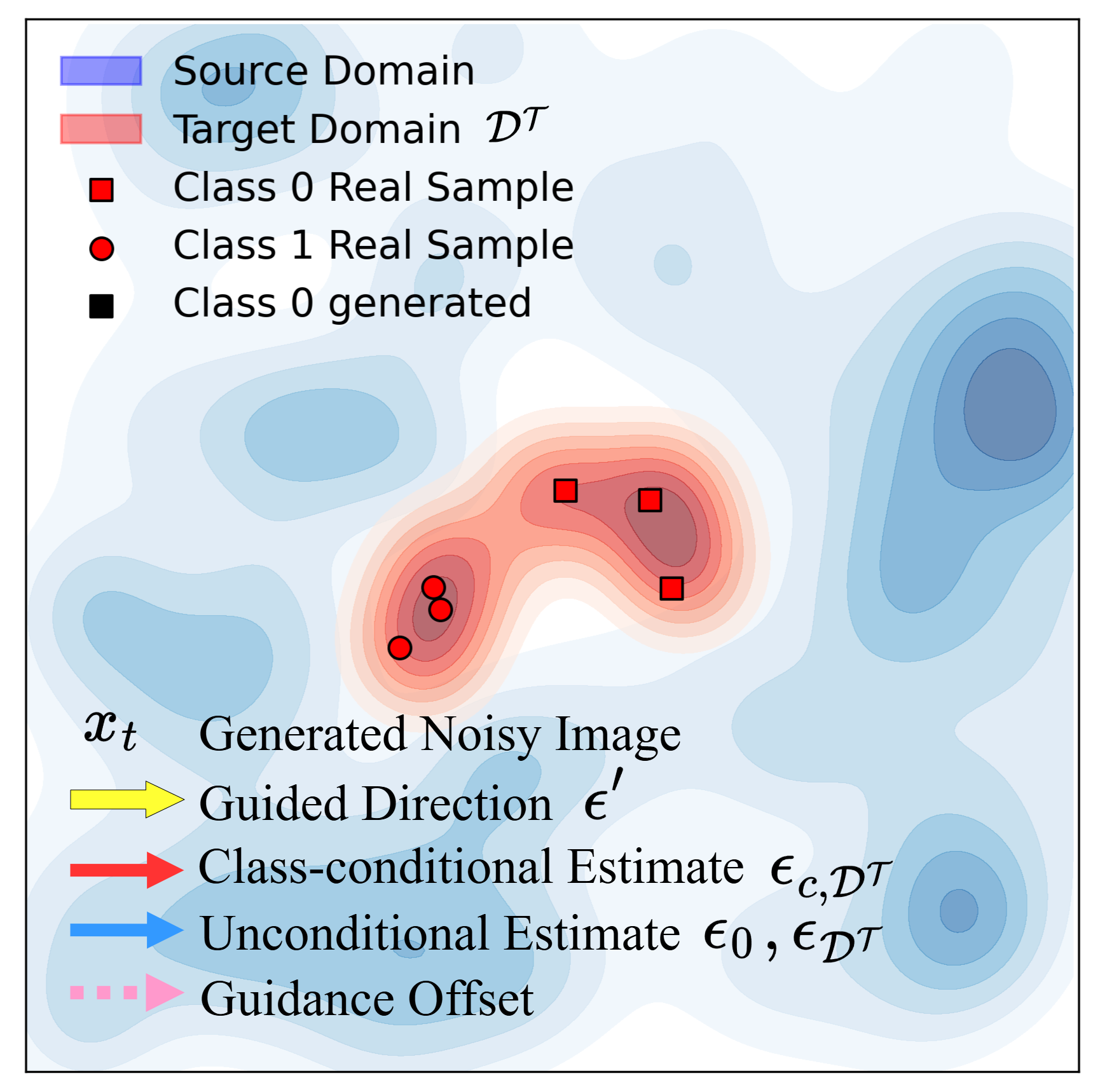}
        \caption{Ground Truth.}
    \end{subfigure}%
    \hfill
    \begin{subfigure}[t]{0.32\textwidth}
        \centering
        \includegraphics[height=5.7cm]{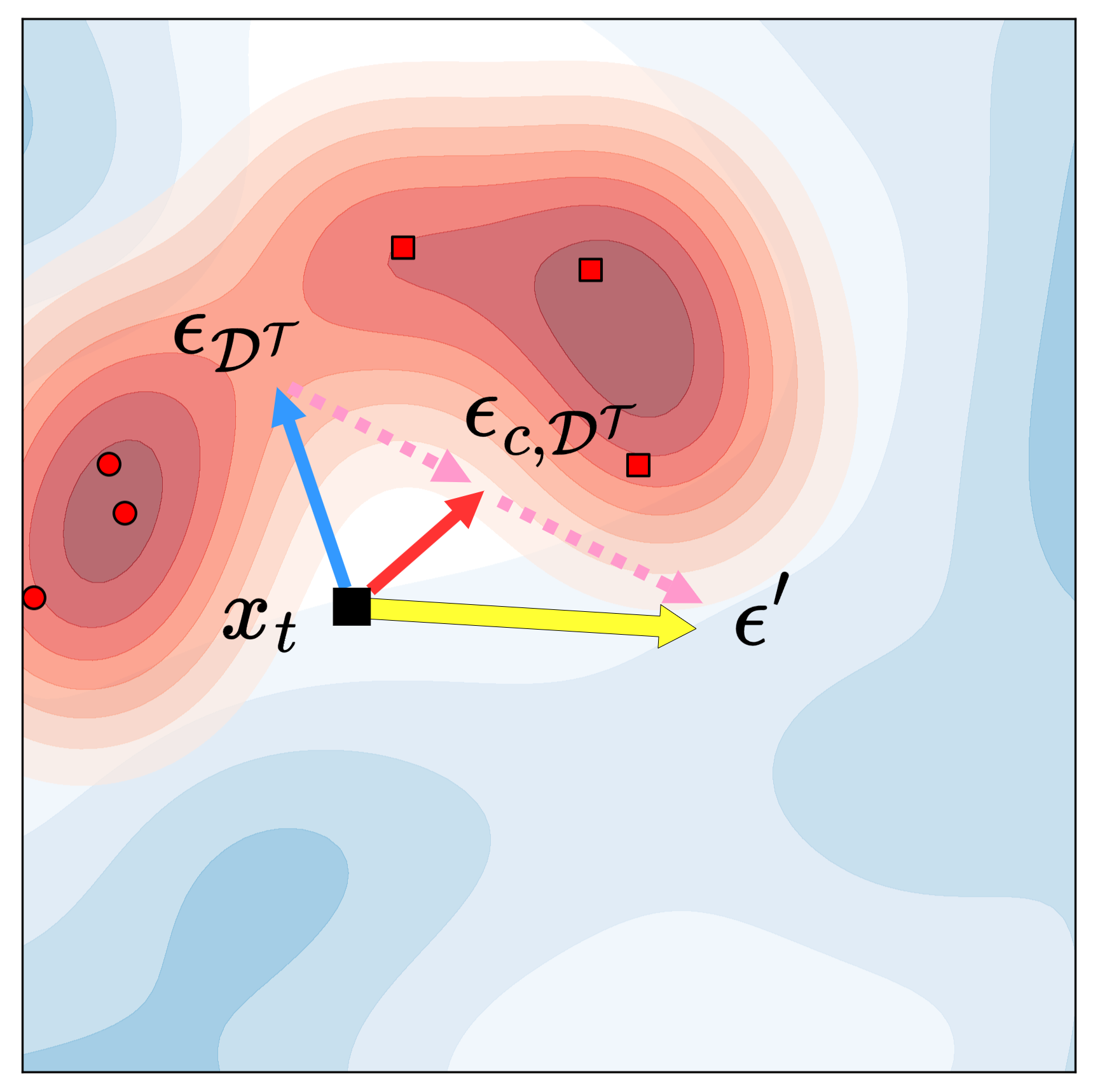}
        \caption{\textcolor{olive}{CFG} and \textcolor{orange}{MG} Directions.}
    \end{subfigure}
    \hfill
    \begin{subfigure}[t]{0.32\textwidth}
        \centering
        \includegraphics[height=5.7cm]{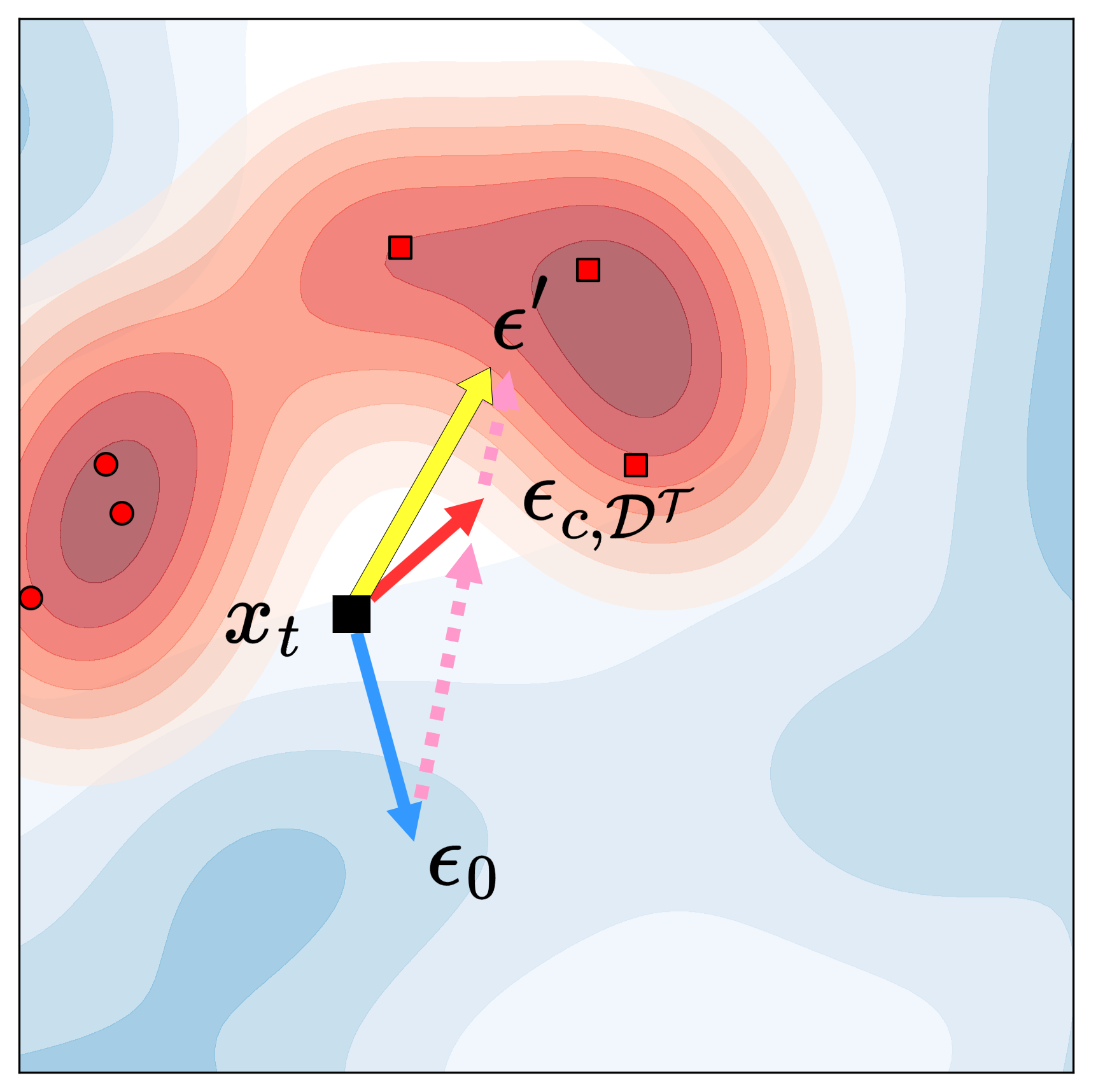}
        \caption{\textcolor{olive}{DoG} and \textcolor{orange}{\textbf{\texttt{DogFit}}} Directions.}
    \end{subfigure}%
    \caption{
    Transfer learning via guidance mechanisms for synthetic data created using a mixture of Gaussian distributions. 
    (a) The red region defines the target domain, while the blue background represents the source distribution.
    (b) CFG and MG prioritize class separability without considering the source distribution, often pushing samples toward out-of-distribution areas in the target domain.
    (c) \textbf{\texttt{DogFit}} and DoG utilize source information to emphasize movement towards the core of the target manifold, improving domain alignment. (*) \textcolor{olive}{Sampling-time} guidance methods (DoG and CFG) operate by computing the \colorbox{pink}{guidance offset}, whereas \textcolor{orange}{training-time} guidance methods (\textbf{\texttt{DogFit}} and MG) learn the \colorbox{yellow}{guided direction} directly. 
    }
    \label{fig:guidance_overview}
\end{figure*}

\noindent\textbf{Diffusion Transfer Learning.}
Transfer learning adapts a model pre-trained on a large-scale source domain to a target domain with smaller size and diversity, leveraging learned representations for better generalization~\citep{weiss2016survey}. In the context of diffusion models, strategies broadly fall into three categories: Parameter-efficient fine-tuning (PEFT) methods reduce training cost by limiting the number of trainable parameters, often via adapters or LoRA~\citep{hu2022lora, xie2023difffit,moon2022fine}. Distillation-based methods preserve pre-trained source priors during adaptation, particularly during early denoising steps~\citep{zhong2024diffusion, hur2024expanding}. Few-shot image generation methods enable diffusion models to generalize to unseen domains using a handful of images~\citep{zhu2022few, wang2024bridging, cao2024few, ouyang2024transfer, ruiz2023dreambooth}. 
Despite this progress, current methods overlook the role of guidance in shaping the generation trajectory, implicitly assuming that its original form remains optimal after adaptation, a limiting assumption in small, low-diversity target domains. We propose a new guidance method tailored for diffusion transfer that can be layered on top of existing fine-tuning methods for improving generalization without incurring additional sampling cost. 
\section{Proposed Method}
\label{sec:methodology}

Let $\epsilon_{\theta_0}$ denote a diffusion model pre-trained on a large-scale source domain $\mathcal{D^S}$, and let $\mathcal{D^T}$ be a smaller target domain such that $|\mathcal{D}^T| \ll |\mathcal{D}^S|$ (typically, $1000<|\mathcal{D}^T|$). The goal of transfer learning is to fine-tune a model $\epsilon_\theta$  (initialized with the weights $\theta_0$ of the pre-trained model) on $\mathcal{D^T}$ such that it generates high-quality, diverse samples aligned with the target distribution $p(x|\mathcal{D^T})$. In this work, we aim to develop a strong guidance mechanism for diffusion transfer learning that is computationally efficient and supports controllable guidance strength.

\subsection{Preliminaries}

\noindent\textbf{Diffusion Models.}
Denoising Diffusion Probabilistic Models (DDPMs)~\citep{ho2020denoising} are generative models that learn to reverse a fixed noising process applied over $T$ time-steps. Starting from a clean image $x_0$, noise is gradually added to produce a sequence of noisy images $\{x_t\}_{t=1}^T$. A neural network $\epsilon_\theta(x_t, t)$ is trained to predict the noise $\epsilon$ at each time-step $t$, using this denoising objective:
\begin{equation}
   \mathcal{L}_{\text{simple}} = \mathbb{E}_{t, x_0, \epsilon} \left[ \| \boldsymbol{\epsilon}_{\theta}(x_t, t) - \boldsymbol{\epsilon} \|^2 \right].
\end{equation}

Time-step $t$ is omitted in subsequent sections when clear from context. In conditional generation, the model receives an auxiliary input $c$ (e.g., class labels or text), producing $\epsilon_\theta(x_t|c)$. SOTA transformer-based diffusion architectures such as DiT~\citep{peebles2023scalable} and SiT~\citep{ma2024sit} operate in latent space rather than directly on pixel space. For notational simplicity, however, we keep using $x$ to denote the model input.

\noindent\textbf{Guidance in Transfer Learning.}
While diffusion models can functionally incorporate conditional inputs during training, their outputs often drift under weak conditioning, motivating the use of guidance signals during the reverse process ~\citep{dhariwal2021diffusion, ho2022classifier}. We present three representative guidance strategies in transfer learning (i.e., CFG, MG, and DoG) covering the main design axes: \emph{where} the guidance signal originates (unconditional branch, or external source model) and \emph{when} it is injected (sampling-time vs.\ training-time). See visual comparison in Fig.~\ref{fig:guidance_overview}.



\begin{figure*}[t]
    \centering

    \begin{minipage}[c]{0.29\textwidth}
        \centering
        \includegraphics[width=\linewidth]{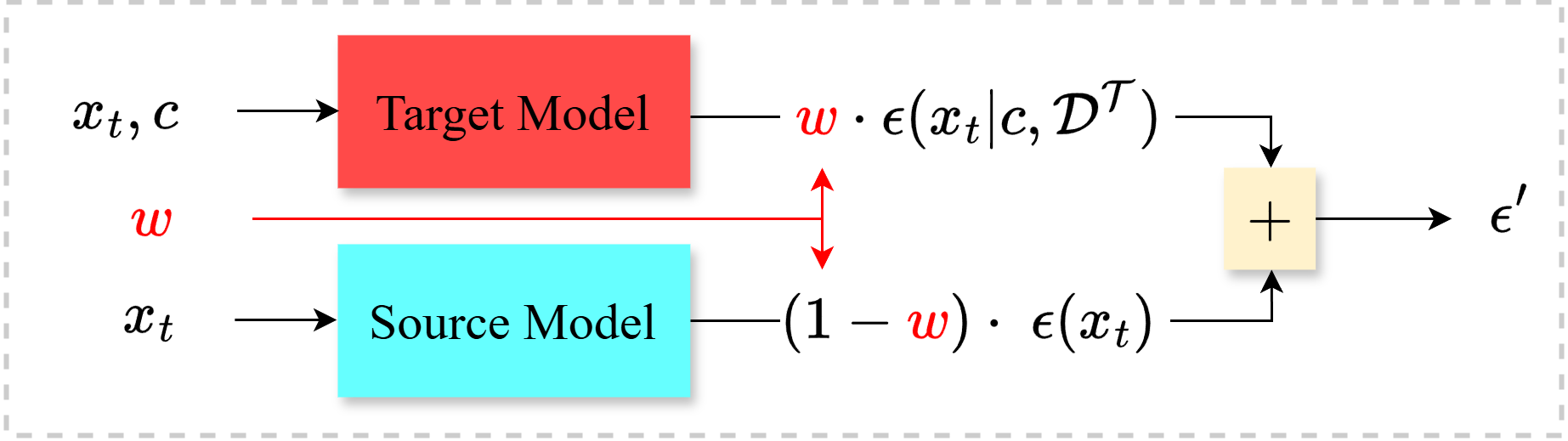}
        \caption*{(a) DoG Sampling.}
        \vspace{1em}
        \includegraphics[width=\linewidth]{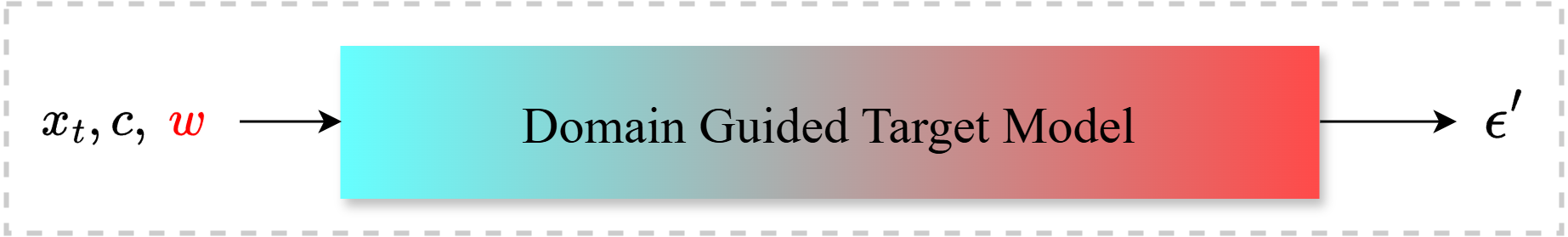}
        \caption*{(b) \texttt{DogFit} Sampling.}
    \end{minipage}%
    \hfill
    \begin{minipage}[c]{0.49\textwidth}
        \vspace*{\fill}  
        \centering
        \includegraphics[height=3.5cm]{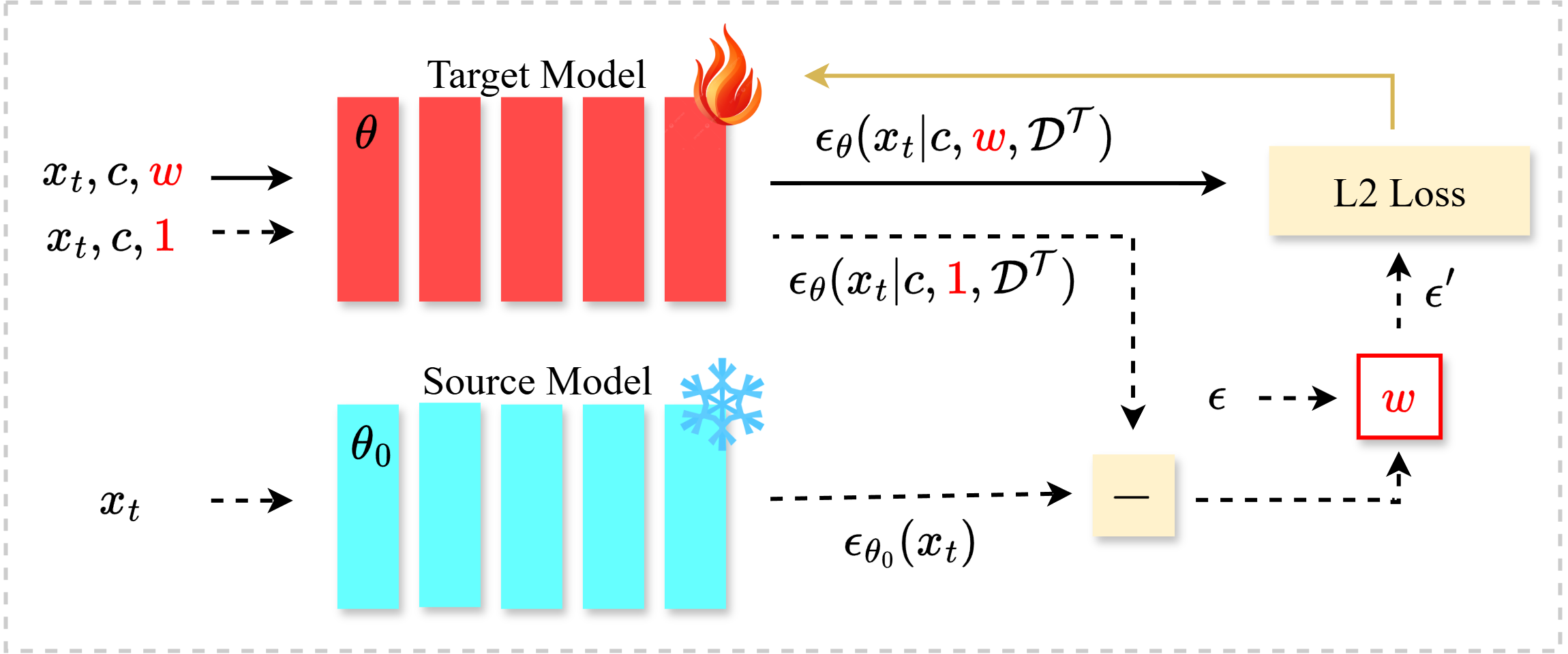}
        \caption*{(c) \texttt{DogFit} Fine-tuning with $w$ Controllability.}
    \end{minipage}
    \begin{minipage}[c]{0.20\textwidth}
        \vspace*{\fill}  
        \centering
        \includegraphics[height=3.5cm]{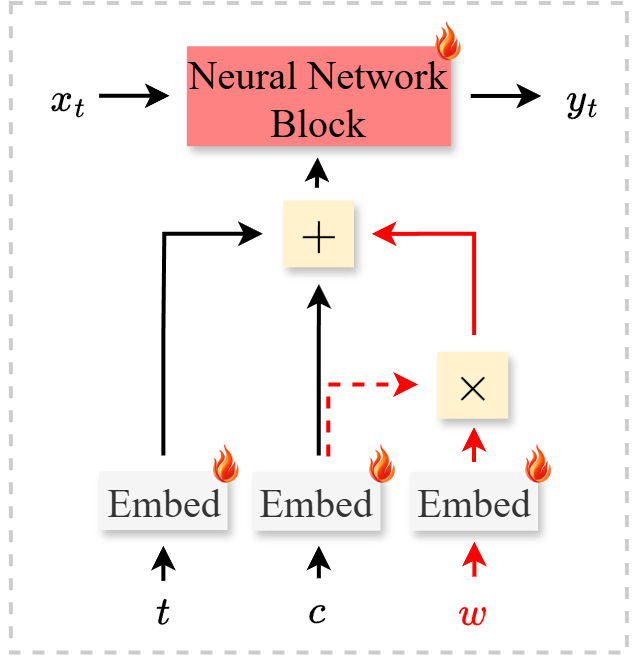}
        \caption*{(d) $w$ Conditioning.}
    \end{minipage}
    \caption{
        Illustration of \texttt{DogFit} during training and sampling
        :(a–b) \texttt{DogFit} internalizes guidance behavior, removing the need for inference-time double forward passes. (c) During training, the guidance value is treated as an input, allowing inference-time control. The simple case of our method, with fixed guidance, simply requires removing the \textcolor{red}{$w$} and \textcolor{red}{$1$} from the target model conditions. (d) \textcolor{red}{$w$} is embedded it as an extra input during fine-tuning, allowing inference-time control. (*) Dashed lines denote paths that do not propagate gradients.
    }
    \label{fig:stacked_left_aligned_right}
\end{figure*}

\noindent\textit{- Classifier-Free Guidance (CFG)}~\citep{ho2022classifier, nichol2021glide} improves conditional generation by extrapolating the model's class-conditional prediction further away from its unconditional counterpart-typically using the same model. The reverse process is modified as:
\begin{equation}
\boldsymbol{\epsilon}'_\theta(x_t|c) = \boldsymbol{\epsilon}_\theta(x_t|c) + (w-1) \cdot \left( \boldsymbol{\epsilon}_\theta(x_t|c) - \boldsymbol{\epsilon}_\theta(x_t) \right),
\end{equation}
where increasing the guidance scale \( w > 1 \) amplifies the effect of the condition during sampling. 
 CFG poses limitations in transfer settings: The unconditional noise estimator is prone to underfitting due to joint training under limited target training data~\citep{chen2023score}. Further, CFG often steers generations toward out-of-distribution regions in the target domain (Fig.~\ref{fig:guidance_overview} (b)), and it relies on conditionally labeled data, which may be unavailable in some domains~\cite{zhong2025domain}.


\noindent\textit{- Domain Guidance (DoG)}~\citep{zhong2025domain} offers a stronger alternative in transfer learning by using the unconditional source model as the marginal noise estimator:
\begin{align}
\boldsymbol{\epsilon}_{\theta}'(x_t| c, \mathcal{D^T}) ={}&\\
\boldsymbol{\epsilon}_{\theta}(x_t \mid c ,\mathcal{D^T}) \notag 
&+ (w -1) \cdot \left( \boldsymbol{\epsilon}_{\theta}(x_t \mid c ,\mathcal{D^T}) -\boldsymbol{\epsilon}_{\theta_0}(x_t) \right),
\end{align}
\noindent where $\boldsymbol{\epsilon}_{\theta}(x_t \mid c, \mathcal{D^T})$ is the fine-tuned target model's class-conditional prediction and $\boldsymbol{\epsilon}_{\theta_0}(x_t)$ is the unconditional one of the source model. The source model, having been trained on rich data, provides a stronger reference point than CFG and further improves target domain alignment~\citep{zhong2025domain} (Fig.~\ref{fig:guidance_overview} (c)). Like CFG, however, DoG requires two forward passes during sampling.

\noindent\textit{- Model Guidance (MG)}~\citep{tang2025diffusion} removes the need for CFG-style computation during sampling by incorporating its effect directly into the training objective. The model is trained to match a guidance-enhanced noise target:
\begin{align}
\label{equation:mg_loss}
\mathcal{L}_{\text{MG}} &= \mathbb{E}_{t, (x_0, c), \boldsymbol{\epsilon}} \left\| \boldsymbol{\epsilon}_\theta(x_t|c) - \boldsymbol{\epsilon}' \right\|^2, \\
\boldsymbol{\epsilon}' &= \boldsymbol{\epsilon} + (w - 1) \cdot \operatorname{sg}\left( \boldsymbol{\epsilon}_\theta(x_t|c) - \boldsymbol{\epsilon}_\theta(x_t) \right), \nonumber
\end{align}
where \( \operatorname{sg}( \cdot) \) denotes the stop-gradient operator and $\epsilon'$ marks the new noise target.
While MG eliminates sampling-time overhead of test-time guidance, it inherits CFG’s limitations in transfer learning due to the underfitting of the unconditional noise estimator. Further, it lacks a clear mechanism for allowing diversity-fidelity control at test time-a crucial principle in diffusion guidance.

\begin{table*}[ht]
\centering
\caption{FID and FD$_\text{DINOV2}$ performance of \texttt{DogFit} against SOTA methods using DiT/XL-2 and SiT/XL-2 backbones. Their sampling costs are also shown in terms of forward passes and TFLOPs.}
\label{tab:fid_fd_dino_combined}
\renewcommand{\arraystretch}{1.3}
\resizebox{\textwidth}{!}{
\begin{tabular}{|c|c|l||c|ccccc|c|cc|}
\hline
& \multirow{2}{*}{\textbf{Metric}} & \multirow{2}{*}{\textbf{Method}}   & \textbf{Unlabeled} & \multicolumn{6}{c|}{\textbf{Labeled}} & \multicolumn{2}{c|}{\textbf{Sampling Cost}} \\
\cline{4-12}
& & & FFHQ & ArtBench & Caltech & CUB-Birds & Food & Stanford-Cars & Avg. & Passes & TFLOPS \\
\hline \hline
\multirow{12}{*}{\rotatebox[origin=c]{90}{\textbf{DiT/XL-2}}}
& \multirow{6}{*}{FID ↓ } 
& Fine-tuning                          & 15.94 & 23.36 & 30.02 &  9.35 & 16.75 & 16.24 & 19.14 & \textbf{x1} & \textbf{366.14} \\
&
& + CFG \cite{ho2022classifier}        &   --  & 20.83 & 24.07 &  5.03 & 11.77 & 10.60 & 14.46 & x2 & 732.28 \\
&
& + DoG \cite{zhong2025domain}         & 13.87 & 17.30 & 23.76 & \textbf{3.65} & 10.97 & \underline{9.77} & 13.09 & x2 & 732.28\\
&
& MG  \cite{tang2025diffusion}         &   --  & 19.91 & 23.71 &  4.85 & 11.13 & 11.03 & 14.13 & \textbf{x1} & \textbf{366.14} \\

&& \texttt{DogFit}     \cellcolor{gray!10}          & \textbf{10.48} & \textbf{16.32} & \textbf{21.68} & \underline{3.69} & \underline{10.64} & \textbf{9.35} & \textbf{12.34} & \textbf{x1} & \textbf{366.14}\\

&
& \texttt{DogFit + Control} \cellcolor{gray!10}  & \underline{13.03} & \underline{16.98} & \underline{21.98} & \underline{3.69} & \textbf{10.44} & {10.05} & \underline{12.63} & \textbf{x1} & \textbf{366.14} \\
\cline{2-12}
&
\multirow{6}{*}{FD$_\text{DINOV2}$ ↓ } 
& Fine-tuning                          & 461.45 & 360.77 & 529.85 & 428.21 & 610.31 & 378.11 & 461.45 & \textbf{x1} & \textbf{366.14} \\
&
& + CFG \cite{ho2022classifier}        &   --   & 314.32 & 401.96 & 200.92 & 410.48 & 227.46 & 311.03 & x2 & 732.28\\
&
& + DoG \cite{zhong2025domain}         & \textbf{273.93} & \underline{266.25} & \underline{377.49} & \textbf{135.74} & {314.16} & \textbf{132.91} & \textbf{245.31} & x2 & 732.28 \\
&
& MG  \cite{tang2025diffusion}         &   --   & 299.06 & 409.20 & 220.40 & 379.39 & 255.88 & 312.78 & \textbf{x1} & \textbf{366.14} \\

&
& \texttt{DogFit}      \cellcolor{gray!10}         & 282.32 & 269.06 & \textbf{377.15} & \underline{143.42} & \textbf{293.24} & {147.20} & \underline{246.01} & \textbf{x1} & \textbf{366.14} \\

 &
& \texttt{DogFit + Control}  \cellcolor{gray!10} &   \underline{274.67}  &  \textbf{261.72}      &   {378.86}     &  {144.39}      &   \underline{314.12}     &    \underline{141.08}    & {248.03}  & \textbf{x1} & \textbf{366.14} \\
\hline
 \hline
 \multirow{12}{*}{\rotatebox[origin=c]{90}{\textbf{SiT/XL-2}}}
 &
 \multirow{6}{*}{FID ↓ } 
& Fine-tuning & \textbf{7.63} & 9.66 & 26.10 & 4.92 & \underline{7.76} & 10.53 & 11.79 & \textbf{x1} & \textbf{454.01} \\
&
& + CFG \cite{ho2022classifier} & - & \underline{9.28} & \underline{23.21} & \underline{3.49} & 7.95 & \underline{9.71} & {10.73} & x2 & 1083.77 \\
&
& + DoG \cite{zhong2025domain} & \textbf{7.63} & 11.22 & 23.53 & \textbf{3.39} & 7.95 & \underline{9.71} & 11.16 & x2 & 1083.77 \\
&
& MG \cite{tang2025diffusion} & - & 9.64 & \textbf{23.10} & 3.60 & \underline{6.84} & \textbf{8.62} & \textbf{10.36} & \textbf{x1} & \textbf{454.01} \\

&
& \texttt{DogFit}\cellcolor{gray!10}  & 12.44 & {9.62} & 23.45 & 3.52 & 7.85 & 10.05 & 10.91 & \textbf{x1} & \textbf{454.01} \\

&
& \texttt{DogFit + Control} \cellcolor{gray!10}  & \underline{10.8} &\textbf{9.03}& 23.15 & 3.59 & \textbf{6.52} & 10.10 & \underline{10.48} & \textbf{x1} & \textbf{454.01} \\
\cline{2-12}
&
\multirow{5}{*}{FD$_\text{DINOV2}$ ↓ }
& Fine-tuning & {335.15} & 271.92 & 519.54 & 405.07 & 522.47 & 283.29 & 400.46 & \textbf{x1} & \textbf{454.01} \\
&
& + CFG \cite{ho2022classifier} & - & 236.13 & 406.63 & 203.47 & 368.26 & 190.24 & 280.94 & x2 & 1083.77 \\
&
& + DoG \cite{zhong2025domain} & {335.15} & \textbf{215.12} & \textbf{387.89} & \underline{160.85} & \underline{298.65} & \textbf{139.09} & \underline{240.32} & x2 & 1083.77 \\
&
& MG \cite{tang2025diffusion} & - & 232.17 & 418.86 & 229.32 & 359.90 & 203.23 & 288.70 & \textbf{x1} & \textbf{454.01} \\

&
& \texttt{DogFit} \cellcolor{gray!10}& \textbf{278.10} & \underline{222.20} & {403.44} & {181.48} & \underline{296.49} & \underline{159.83} & {252.29} & \textbf{x1} & \textbf{454.01} \\

&
& \texttt{DogFit + Control} \cellcolor{gray!10}   & \underline{319.16} & 230.23 & \underline{399.20} & \textbf{152.50} & \textbf{234.98} & 183.41 & \textbf{240.06} & \textbf{x1} & \textbf{454.01} \\ \hline
\end{tabular}
}
\end{table*}

\subsection{Domain Guided Fine-Tuning}


We propose \texttt{DogFit}, a training-time guidance mechanism for diffusion transfer learning that integrates a domain-aware guidance direction directly into the loss during fine-tuning. The marginal estimate of \texttt{DogFit} is derived from the unconditional source model (Fig.\ref{fig:stacked_left_aligned_right} (a-b)). It modifies the training loss as:
\begin{align}
\label{eq:main_eq}
\mathcal{L}_{\text{\texttt{DogFit}}} &= \mathbb{E}_{t, (x_0, c), \epsilon} \left\| \boldsymbol{\epsilon}_\theta(x_t |c, \mathcal{D^T}) - \boldsymbol{\epsilon}' \right\|^2, \\
\boldsymbol{\epsilon}' &= \boldsymbol{\epsilon} + (w-1) \cdot \text{sg}\left( \boldsymbol{\epsilon}_\theta(x_t | c, \mathcal{D^T}) - \boldsymbol{\epsilon}_{\theta_0}(x_t) \right). \nonumber
\end{align}

The training architecture is illustrated in Fig.~\ref{fig:stacked_left_aligned_right}(c). The training algorithm is shown in Appx.~\ref{app:algorithm}.  The rest of this subsection provides theoretical insights motivating the design of \texttt{DogFit} (full statements and proofs in Appx.~\ref{app:theory}). 

\paragraph{Proposition 1.}
\textit{Suppose a model is trained using the \texttt{DogFit} objective with controllable guidance strength $w$ (Eq.~\ref{eq:main_control}), and the following assumptions hold: (i) the model approximately recovers the true noise when no guidance is applied, and (ii) its predictions vary linearly with $w$. Then, for any value of $w$, the model replicates the effect of applying DoG at sampling time.}

\paragraph{Proposition 2.}
\textit{\texttt{DogFit} is equivalent to implicitly applying a domain alignment component to MG's training loss, benefiting from the class-conditional guidance of MG while reducing the risk of out-of-distribution generation.}

\paragraph{}
This means that $\epsilon_{\theta_0}$ provides a strong and stable marginal noise estimate not only as an inference-time signal, but also during fine-tuning. Further, the domain alignment term, independent of class conditioning, pulls samples toward the core of the target data manifold, rendering \texttt{DogFit} effective even in the unconditional setting. In summary, it combines the strengths of previous guidance mechanisms in transfer learning into a single framework: it inherits the fast, single-pass sampling of MG and the robust marginal noise estimator of DoG. W



\subsection{Incorporating Controllability}
\label{sec:controllable_guidance}
\begin{figure*}[t]
    \centering
        \includegraphics[width=\linewidth]{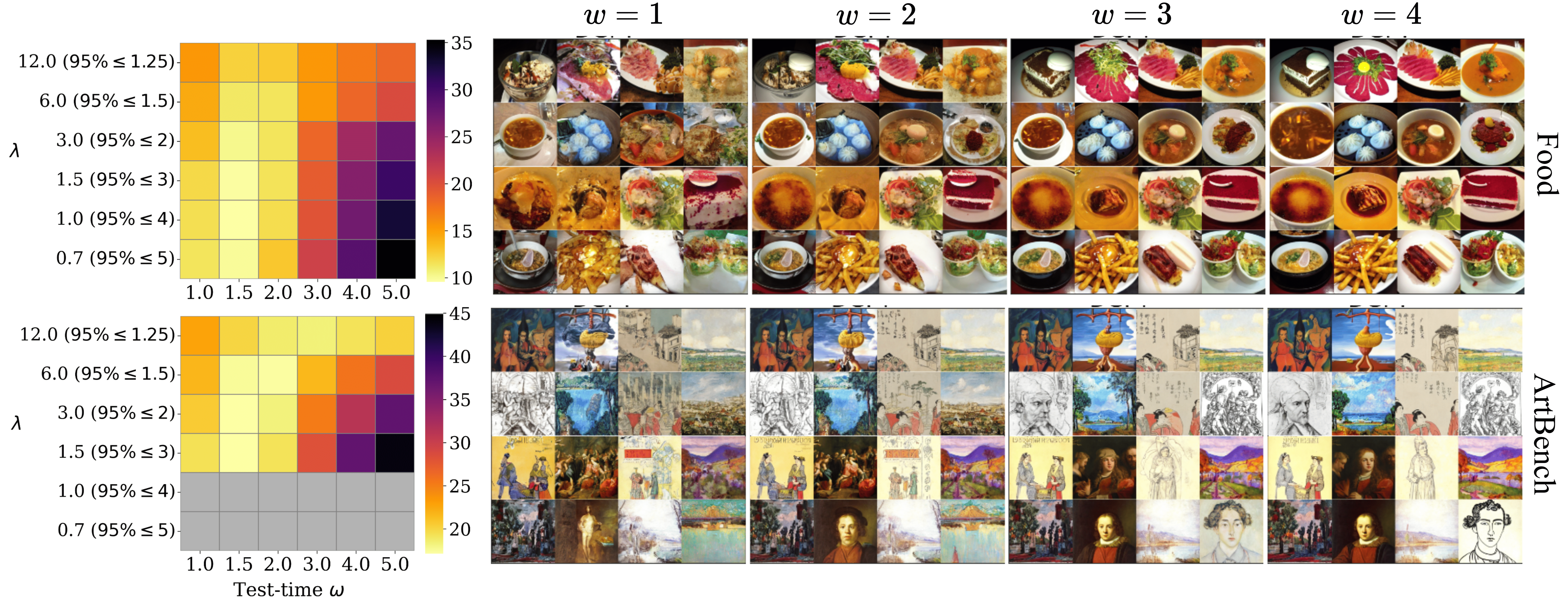} \\
    \caption{Effect of controllable guidance at test time for the Food and ArtBench domains. (Left) FID behavior across test-time $w$s with different training-time $\lambda$s. (Right) Corresponding generated samples for fixed $\lambda = 3$ (95\% of sampled $w$ values in $[1,2]$), across varying test-time $w$. \colorbox{lightgray}{Gray cells} indicate extreme FID values ($\approx$350) resulting from collapsed generations.}
\label{fig:controllable_guidance_plot}
\end{figure*}

A crucial feature of guidance is its controllability. To enable efficient fidelity–diversity trade-offs at test time, we introduce a lightweight conditioning mechanism to the model and 
treat $w$ as an additional input condition during fine-tuning. The training objective with added control is:
\begin{align}
\label{eq:main_control}
\mathcal{L}_{\text{\texttt{DogFit}}} &= \mathbb{E}_{t, (x_0, c, \boldsymbol{w}), \epsilon} \left\| \epsilon_\theta(x_t |c,  \boldsymbol{w},  \mathcal{D^T}) - \epsilon' \right\|^2, \\
\epsilon' &= \epsilon + (\boldsymbol{w}-1) \cdot \text{sg}\left( \epsilon_\theta(x_t |c,  \boldsymbol{1}, \mathcal{D^T}) - \epsilon_{\theta_0}(x_t) \right),\nonumber
\end{align}
\noindent where \( \epsilon_\theta(x_t \mid c, \boldsymbol{1}, \mathcal{D^T}) \) is the unguided class-conditional prediction. During training, the model must learn to operate across a range of guidance strengths to support controllable fidelity–diversity trade-offs at inference, while still preserving stable behavior in the unguided setting. Our initial experiments revealed that a simple uniform sampling of \( \boldsymbol{w} \) during training can reduce diversity and destabilize generation, even at lower guidance levels due to excessive exposure to high $w$. We sample \( \boldsymbol{w} \) from a shifted exponential decaying distribution (SEDD; see distribution CDF in Appx.~\ref{app:theory}):
\begin{align}
\boldsymbol{w} &= 1 + z, \quad z \sim \mathcal{P}(z), \\
\mathcal{P}(z) &= \lambda e^{-\lambda z}, \quad z \geq 0,
\label{eq:exponential} 
\end{align}
\noindent where \( \lambda \) controls the decay rate. 
This distribution favors smaller guidance strengths while occasionally introducing larger values, enabling guidance robustness without destabilizing training. As a result, the model generalizes well across the guidance spectrum and supports dynamic controllability at inference (Fig.~\ref{fig:controllable_guidance_plot}; see comparison with uniform scheduling in Appx.~\ref{app:theory})

The next key question is \textit{how to actually incorporate $w$ into the model?}
\begin{figure}[t]
    \centering
    \includegraphics[width=\linewidth]{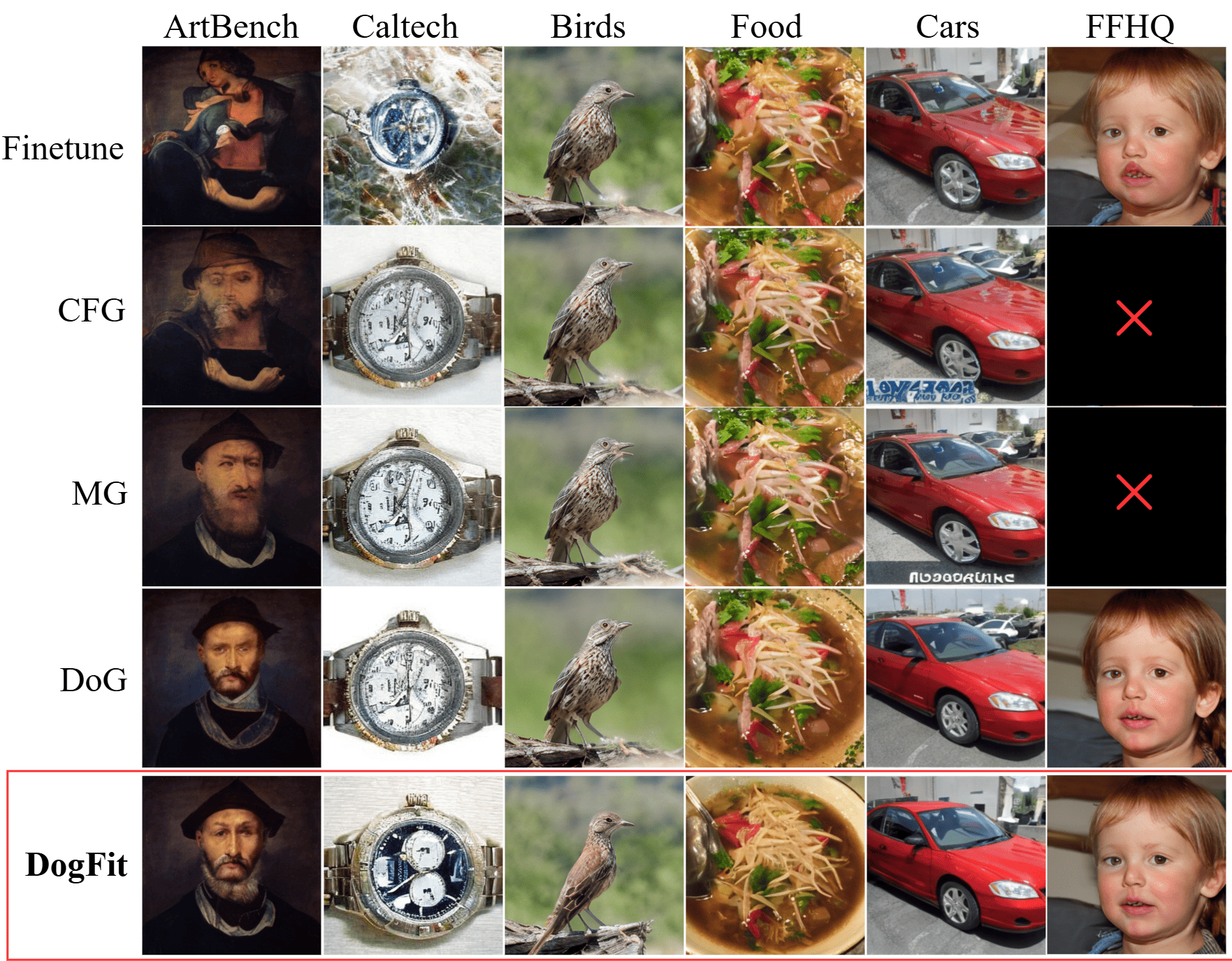}
    \caption{Qualitative comparison of guidance mechanisms on DiT-XL/2 (guidance scale 1.5). $\textcolor{red}{\times}$: not applicable.}
    \label{fig:qualitative_comparison}
\end{figure}
In conditional diffusion models, the final condition embedding is typically formed by summing the label $c$ and time-step $t$ embeddings. Therefore, a natural approach to add $w$ is to embed it and sum it with the other existing embeddings. However, we saw that even slight distributional shifts and directional changes can destabilize training and disrupt the pretrained features. To avoid this, we use $w$ as a label modulator: we embed it as a scalar, multiply it with a detached label $c$ embedding, and add the resulting modulation vector to the condition embedding. This allows $w$ to control the influence of the label without altering its semantic direction, preserving the integrity of the pretrained embedding space and ensuring stable guidance conditioning. For the unconditional case, we use the same formulation while keeping $c$ as a fixed zero vector. 

\begin{figure*}[t]
    \centering

    \begin{minipage}[t]{0.49\textwidth}
        \centering
        \includegraphics[width=\linewidth]
        {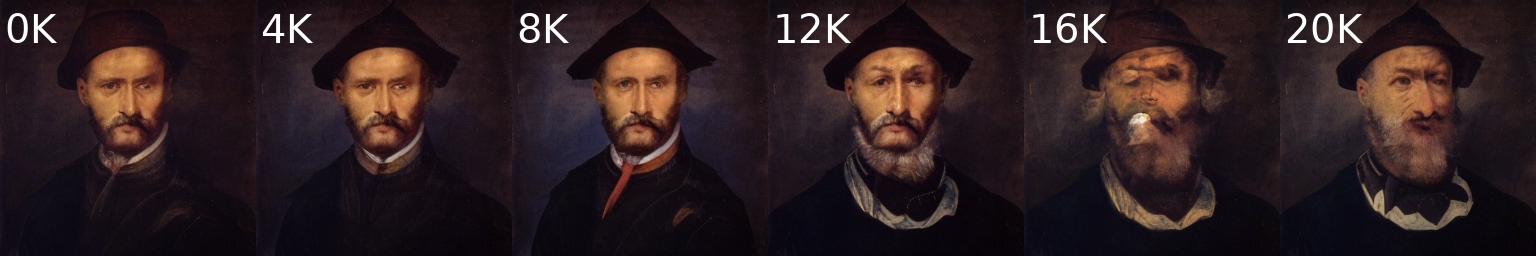} \\[4pt]
        \includegraphics[width=\linewidth]{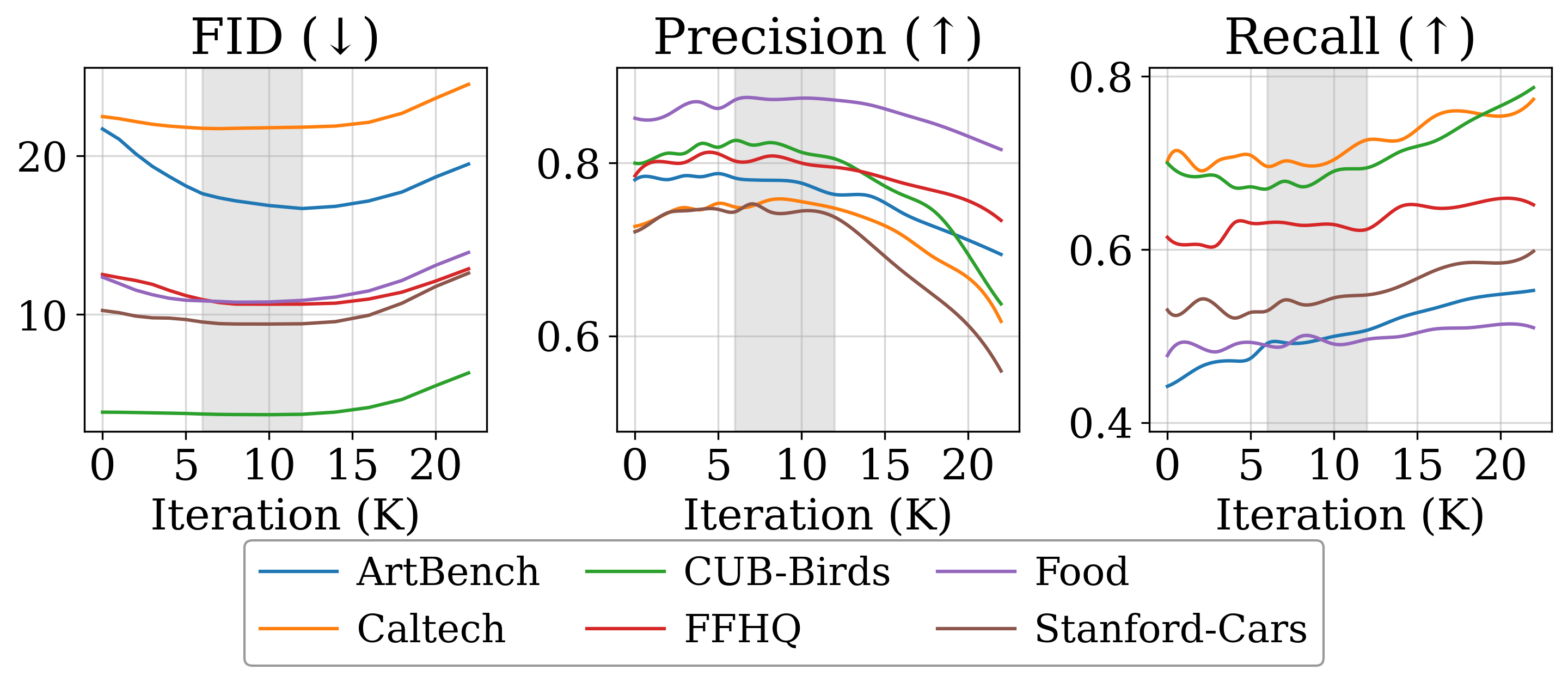}
        \textbf{(a)} Late-start.
    \end{minipage}
    \hfill
    \begin{minipage}[t]{0.49\textwidth}
        \centering
        \includegraphics[width=\linewidth]
        {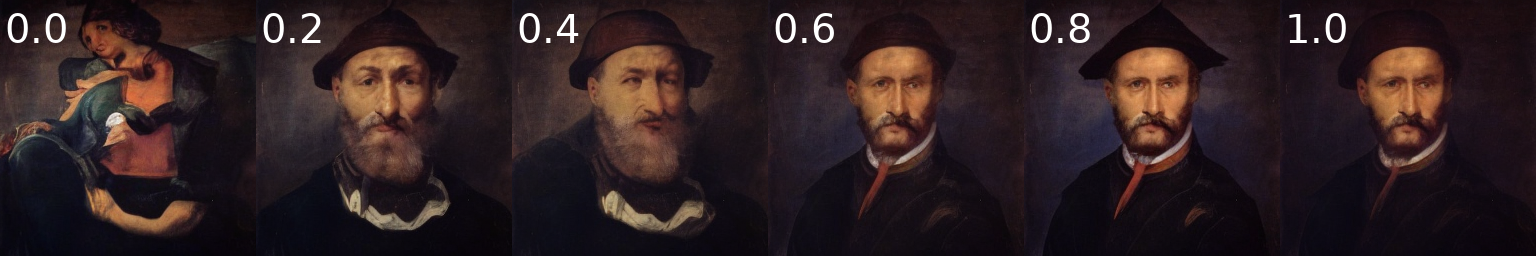} \\[4pt]
        \includegraphics[width=\linewidth]{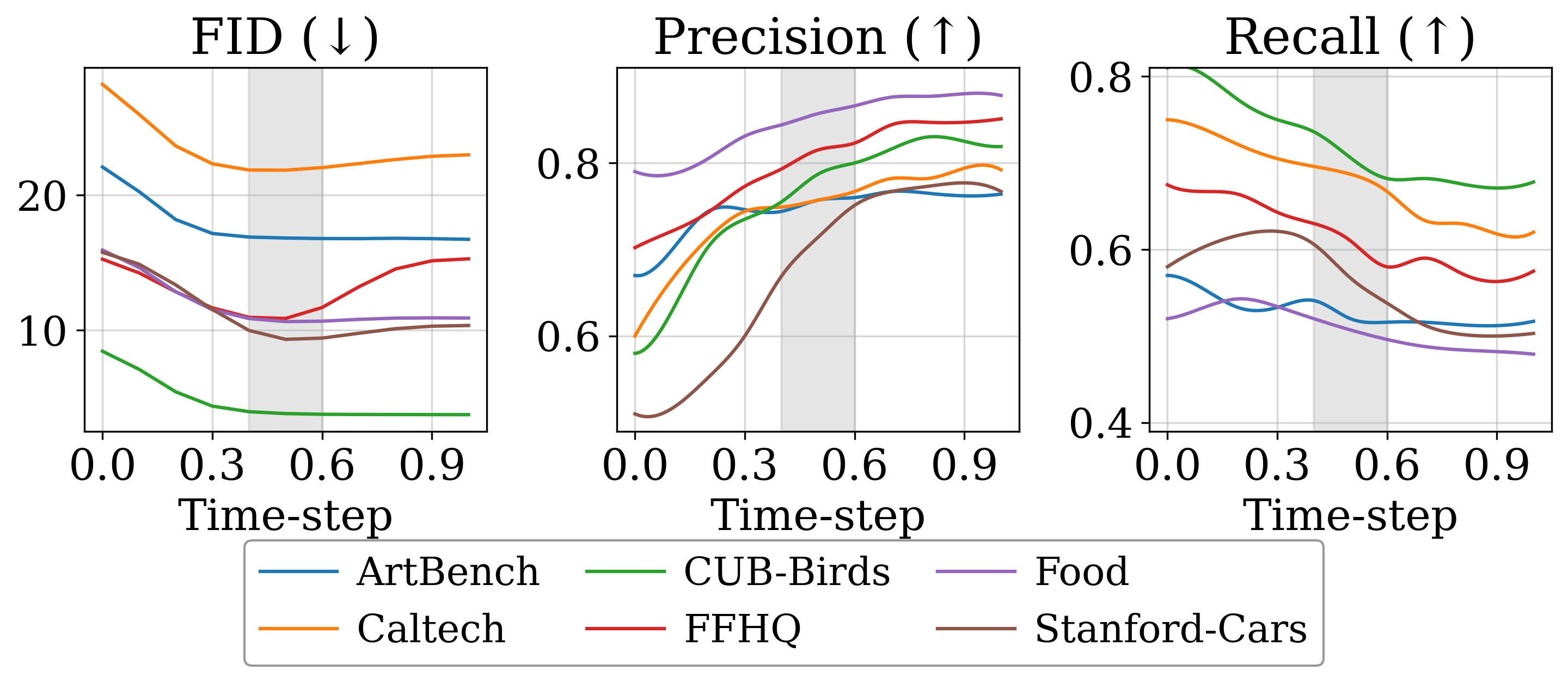}
        \vspace{0.5em}
        \textbf{(b)} Cut-off.
    \end{minipage}
    \caption{Ablation study on guidance schedules in \texttt{DogFit}. (a) Varying the late-start threshold $\tau_{\text{s}}$ to control when guidance begins. (b) Varying the cut-off threshold $\tau_{\text{c}}$ to restrict guidance to later denoising steps. Performed using FID on DiT-XL/2.}
    \label{fig:ablation_grid}
\end{figure*}

\subsection{Guidance Scheduling Mechanisms}
Inspired by test-time guidance scheduling for improved generation quality~\citep{sadat2023cads, kynkaanniemi2024applying}, we investigate the optimal timing and placement of training-time guidance using two simple yet effective strategies. First, we introduce a \textit{late-start} mechanism that delays the injection of guidance until iteration $s > \tau_{\text{s}}$, avoiding noisy updates and unstable gradients in the early phases of training. This is particularly beneficial in conditional settings where class embeddings are reinitialized to accommodate mismatched source and target label spaces. By deferring guidance, the model benefits from more stable target representations and condition embeddings, resulting in higher-fidelity updates. 
Second, we introduce a \textit{cut-off} strategy that disables guidance at higher noise levels ($t > \tau_{\text{c}}$), concentrating domain supervision on the later denoising stages where adaptation to target-specific details is most effective~\citep{choi2022perception}. This design preserves source-driven diversity in the early denoising steps and prevents the model from over-relying on the target domain to shape the global structure of the image. This becomes crucial when source and target domains share similar structural priors (e.g. FFHQ faces).

\section{Results and Discussion}
\label{sec:setup}

\noindent\textbf{Setup.} 
\texttt{DogFit} is validated on six challenging datasets with diverse domain characteristics and label availability for a comprehensive assessment: Food101~\citep{bossard2014food}, Caltech101~\citep{griffin2007caltech}, CUB-200-201~\citep{wah2011caltech}, ArtBench10~\citep{liao2022artbench}, Stanford-Cars~\citep{krause20133d}, and FFHQ256~\citep{karras2019style}. These datasets range from fine-grained species~\citep{wah2011caltech} to abstract art-styles~\citep{liao2022artbench} and face generation~\citep{karras2019style}, covering a variety of distribution shifts in appearance, texture, and structure. We compare \texttt{DogFit} to standard fine-tuning, CFG, DoG, and MG. All methods are applied on DiT-XL/2~\citep{peebles2023scalable} and SiT-XL/2~\citep{ma2024sit} backbones pre-trained on ImageNet at 256$\times$256 resolution for 7M steps with Fréchet Inception Distance (FID)~\citep{heusel2017gans} scores of 2.27 (DiT) and 2.06 (SiT).
We generate 10,000 images using 50 sampling steps~\cite{peebles2023scalable, xie2023difffit}, fixing guidance strength to 1.5 for fair comparison, and fine-tune for 24,000 training steps ~\cite{zhong2025domain} with batch size of 32 on 2 A100 GPUs for 5 hours. We calculate FID and $\text{FD}_{\text{DINOV2}}$~\citep{stein2023exposing} between the generated images and the full datasets to evaluate the generation quality. Further, we compute Precision and Recall \citep{kynkaanniemi2019improved} for analyzing fidelity and diversity. 

\subsection{Comparison with State-of-the-art Methods}

Tab.\ref{tab:fid_fd_dino_combined} summarizes our main quantitative findings. \texttt{DogFit} consistently performs comparatively or superior to prior methods in both FID and $\text{FD}_{\text{DINOV2}}$, while maintaining the efficiency of one-pass sampling. Unlike CFG and MG, \texttt{DogFit} remains effective in unconditional settings like FFHQ due to its class-agnostic domain-alignment direction. On SiT, CFG and MG slightly outperform \texttt{DogFit} in FID, but \texttt{DogFit} outperforms both CFG and MG on $\text{FD}_{\text{DINOV2}}$, underscoring the value of utilizing diverse metrics for comprehensive evaluation. Importantly, equipping \texttt{DogFit} with controllable guidance introduces no performance degradation while increasing the parameter count by only 1.33 million parameters (less than 2\% of the initial network parameters), leaving no notable impact on sampling TFLOPS. Additional results on Precision and Recall are provided in Appx.~\ref{app:exp_precision_recall}. Fig.~\ref{fig:qualitative_comparison} presents a visual comparison of generated samples across target domains on DiT. Guidance-based methods consistently outperform naïve fine-tuning by producing more realistic and coherent images. \texttt{DogFit} generates sharp, domain-relevant details on par with or better looking than DoG, while avoiding off-distribution artifacts occasionally seen with CFG and MG (e.g., distorted faces or tires). SiT qualitative results are provided in Appx.~\ref{app:exp_sit}.


\subsection{Ablation Studies}
\noindent\textbf{Controllable Guidance}
We study the effect of $\lambda$ for sampling the guidance scalar $w$ from the SED during training (Eq.~\ref{eq:exponential}). As shown in Fig.~\ref{fig:controllable_guidance_plot} (left), sparse exposure to higher guidance is sufficient for generalization and achieving a diversity-fidelity tradeoff. In contrast, excessive exposure to large guidance can harm the model’s ability to retain pretrained features, leading to instability and collapsed generations. Based on these findings, we adopt $\lambda = 3$ (95\% of sampled $w$s in $[1,2]$) for all experiments. Fig.~\ref{fig:controllable_guidance_plot} (right) illustrates how varying guidance strength at test time effectively modulates the trade-off between fidelity and diversity, increasing sample quality while reducing variation with higher guidance values (more analysis in Appx.~\ref{app:exp_sensitivity}).

\noindent \textbf{Guidance Scheduling.} 
We analyze the impact of late-start and cut-off on generation quality and diversity. Fig.~\ref{fig:ablation_grid}(a) shows how varying the late-start threshold $\tau_s$ impacts FID, Precision, and Recall. Delaying guidance improves FID and especially Precision at first, as early stabilization allows formation of more accurate directional cues. However, setting $\tau_s$ too high leaves insufficient time for the model to internalize guidance, reducing it to naïve fine-tuning and lowering precision. We set $\tau_s = 12\text{k}$ for FID and $\tau_s = 6\text{k}$ for $\text{FD}_{\text{DINOV2}}$ in all experiments. 
Fig.~\ref{fig:ablation_grid}(b) shows the effect of cut-off threshold $\tau_c$ on FID, Precision, and Recall. Delaying guidance to later denoising steps enhances diversity, particularly for target datasets with strong global structural similarity to the source domain, such as FFHQ, Cars, and Caltech. However, setting $\tau_c$ too low limits exposure to guidance, again reducing the model to naïve fine-tuning. While the optimal $\tau_c$ may depend on the source–target similarity, we use $\tau_c = 0.5$ for FID and $\tau_c = 1$ for $\text{FD}_{\text{DINOV2}}$ in our experiments ($t\in [0,1]$). See $\text{FD}_{\text{DINOV2}}$ analysis in Appx.~\ref{app:fd_dino}.

\begin{table}[t]
\centering
\scriptsize
\begin{tabular}{|c|cc|cc|}
\hline
\textbf{\multirow{2}{*}{\textbf{Steps}}} & \multicolumn{2}{c|}{\textbf{Food }\textbf{FID↓}} & \multicolumn{2}{c|}{\textbf{Art} \textbf{FID↓}} \\
\cline{2-5}
               & {MG} & {\texttt{DogFit}} & {MG} & {\texttt{DogFit}} \\
\hline\hline
10  & 60.60  & \textbf{57.91} & 108.26 & \textbf{88.84} \\
25  & 21.13  & \textbf{18.97} & 39.80  & \textbf{29.11} \\
50  & 11.13  & \textbf{10.64} & 19.91  & \textbf{16.32} \\
100 & \textbf{7.60}   & 8.09 & 13.38  & \textbf{12.70} \\
\hline
\end{tabular}
\caption{Sampling steps for MG and \texttt{DogFit} methods using DiT-XL/2.}
\label{tab:fid_vs_steps}
\end{table}

\begin{table}[t]
\centering
\scriptsize
\centering
\begin{tabular}{|l | cc | c|}
\hline
\textbf{DiT-XL/2} & \multicolumn{2}{c|}{\textbf{FID↓}} & \textbf{Train}  \\
\cline{2-3}
\textbf{Variant} & Food & Art & \textbf{Params (M)}  \\
  \hline\hline
\texttt{DogFit}              & \textbf{10.64} & 16.32 & 675.42 (100\%) \\
+ DiffFit           & 11.86 & \textbf{15.80} & \textbf{0.75 (0.11\%)} \\
\hline
\end{tabular}
\caption{Combining \texttt{DogFit} with DiffFit.}
\label{tab:dit-difffit}
\end{table}

\noindent \textbf{Domain-guided Parameter-efficient Fine-tuning.}
We demonstrate that \texttt{DogFit} is fully compatible with DiffFit, a leading PEFT method for DiT models~\cite{xie2023difffit}. As shown in Tab.\ref{tab:dit-difffit}, combining the two yields high-quality generations with lower cost in both training and test-time. This highlights \texttt{DogFit}’s practicality as a simple drop-in enhancement for efficient fine-tuning mechanisms.

\noindent \textbf{Sampling Steps.} Tab.\ref{tab:fid_vs_steps} reports the effect of varying the number of sampling steps during generation. \texttt{DogFit} outperforms MG in nearly all settings, with especially notable gains at lower step counts. This underscores the effectiveness of \texttt{DogFit}’s target alignment in low-resource scenarios where fast sampling is critical.



\section{Conclusion}

We introduce \texttt{DogFit}, a training-time guidance strategy for diffusion model transfer learning. It eliminates the high sampling-time cost of test-time guidance while preserving strong generalization behavior and controllability. \texttt{DogFit} leverages the pre-trained source model to inject high quality, domain-aligned guidance signals directly into the fine-tuning loss, and relies on a lightweight conditioning mechanism for inference fidelity–diversity adjustments. Our training-time guidance scheduling strategies further enhance training stability and generation quality. Extensive experiments across diverse target datasets and models show that \texttt{DogFit} can surpass the performance and efficiency of SOTA guidance methods, establishing it as a practical solution for scalable diffusion model transfer.

\noindent\textbf{Limitations and Future Work.} While effective, the proposed guidance scheduling strategies rely on manually chosen, fixed thresholds. Future work explores adaptive or data-driven approaches that respond to training dynamics or domain-specific properties. Moreover, extension of \texttt{DogFit} it to broader settings, i.e., text-to-image, audio, and video generation remains an exciting direction.

\noindent\textbf{Acknowledgments.} This research was supported by the Natural Sciences and Engineering Research Council of Canada, and the Digital Research Alliance of Canada.

\bibliography{aaai2026}

@String(ICLR = {Int. Conf. Learn. Represent.})

@String(ICLR  = {ICLR})

@article{ho2020denoising,
  title={Denoising diffusion probabilistic models},
  author={Ho, Jonathan and Jain, Ajay and Abbeel, Pieter},
  journal={Advances in neural information processing systems},
  volume={33},
  pages={6840--6851},
  year={2020}
}

@inproceedings{sohl2015deep,
  title={Deep unsupervised learning using nonequilibrium thermodynamics},
  author={Sohl-Dickstein, Jascha and Weiss, Eric and Maheswaranathan, Niru and Ganguli, Surya},
  booktitle={International conference on machine learning},
  pages={2256--2265},
  year={2015},
  organization={pmlr}
}

@article{song2020denoising,
  title={Denoising diffusion implicit models},
  author={Song, Jiaming and Meng, Chenlin and Ermon, Stefano},
  journal={arXiv preprint arXiv:2010.02502},
  year={2020}
}

@inproceedings{rombach2022high,
  title={High-resolution image synthesis with latent diffusion models},
  author={Rombach, Robin and Blattmann, Andreas and Lorenz, Dominik and Esser, Patrick and Ommer, Bj{\"o}rn},
  booktitle={Proceedings of the IEEE/CVF conference on computer vision and pattern recognition},
  pages={10684--10695},
  year={2022}
}

@article{podell2023sdxl,
  title={Sdxl: Improving latent diffusion models for high-resolution image synthesis},
  author={Podell, Dustin and English, Zion and Lacey, Kyle and Blattmann, Andreas and Dockhorn, Tim and M{\"u}ller, Jonas and Penna, Joe and Rombach, Robin},
  journal={arXiv preprint arXiv:2307.01952},
  year={2023}
}

@inproceedings{gupta2024photorealistic,
  title={Photorealistic video generation with diffusion models},
  author={Gupta, Agrim and Yu, Lijun and Sohn, Kihyuk and Gu, Xiuye and Hahn, Meera and Li, Fei-Fei and Essa, Irfan and Jiang, Lu and Lezama, Jos{\'e}},
  booktitle={European Conference on Computer Vision},
  pages={393--411},
  year={2024},
  organization={Springer}
}

@article{ho2022video,
  title={Video diffusion models},
  author={Ho, Jonathan and Salimans, Tim and Gritsenko, Alexey and Chan, William and Norouzi, Mohammad and Fleet, David J},
  journal={Advances in Neural Information Processing Systems},
  volume={35},
  pages={8633--8646},
  year={2022}
}

@inproceedings{kawar2023imagic,
  title={Imagic: Text-based real image editing with diffusion models},
  author={Kawar, Bahjat and Zada, Shiran and Lang, Oran and Tov, Omer and Chang, Huiwen and Dekel, Tali and Mosseri, Inbar and Irani, Michal},
  booktitle={Proceedings of the IEEE/CVF conference on computer vision and pattern recognition},
  pages={6007--6017},
  year={2023}
}

@article{dhariwal2021diffusion, 
  title={Diffusion models beat gans on image synthesis},
  author={Dhariwal, Prafulla and Nichol, Alexander},
  journal={Advances in neural information processing systems},
  volume={34},
  pages={8780--8794},
  year={2021}
}

@article{ho2022classifier,
  title={Classifier-free diffusion guidance},
  author={Ho, Jonathan and Salimans, Tim},
  journal={arXiv preprint arXiv:2207.12598},
  year={2022}
}

@article{nichol2021glide, 
  title={Glide: Towards photorealistic image generation and editing with text-guided diffusion models},
  author={Nichol, Alex and Dhariwal, Prafulla and Ramesh, Aditya and Shyam, Pranav and Mishkin, Pamela and McGrew, Bob and Sutskever, Ilya and Chen, Mark},
  journal={arXiv preprint arXiv:2112.10741},
  year={2021}
}

@article{tang2025diffusion,
  title={Diffusion Models without Classifier-free Guidance},
  author={Tang, Zhicong and Bao, Jianmin and Chen, Dong and Guo, Baining},
  journal={arXiv preprint arXiv:2502.12154},
  year={2025}
}

@article{chen2025visual,
  title={Visual Generation Without Guidance},
  author={Chen, Huayu and Jiang, Kai and Zheng, Kaiwen and Chen, Jianfei and Su, Hang and Zhu, Jun},
  journal={arXiv preprint arXiv:2501.15420},
  year={2025}
}

@article{kynkaanniemi2024applying,
  title={Applying guidance in a limited interval improves sample and distribution quality in diffusion models},
  author={Kynk{\"a}{\"a}nniemi, Tuomas and Aittala, Miika and Karras, Tero and Laine, Samuli and Aila, Timo and Lehtinen, Jaakko},
  journal={arXiv preprint arXiv:2404.07724},
  year={2024}
}

@article{karras2024guiding,
  title={Guiding a diffusion model with a bad version of itself},
  author={Karras, Tero and Aittala, Miika and Kynk{\"a}{\"a}nniemi, Tuomas and Lehtinen, Jaakko and Aila, Timo and Laine, Samuli},
  journal={Advances in Neural Information Processing Systems},
  volume={37},
  pages={52996--53021},
  year={2024}
}

@article{ouyang2024transfer,
  title={Transfer Learning for Diffusion Models},
  author={Ouyang, Yidong and Xie, Liyan and Zha, Hongyuan and Cheng, Guang},
  journal={arXiv preprint arXiv:2405.16876},
  year={2024}
}

@inproceedings{wang2024bridging,
  title={Bridging data gaps in diffusion models with adversarial noise-based transfer learning},
  author={Wang, Xiyu and Lin, Baijiong and Liu, Daochang and Chen, Ying-Cong and Xu, Chang},
  booktitle={Forty-first International Conference on Machine Learning},
  year={2024}
}

@article{zhong2025domain,
  title={Domain guidance: A simple transfer approach for a pre-trained diffusion model},
  author={Zhong, Jincheng and Zhang, Xiangcheng and Wang, Jianmin and Long, Mingsheng},
  journal={arXiv preprint arXiv:2504.01521},
  year={2025}
}

@article{phunyaphibarn2025unconditional,
  title={Unconditional Priors Matter! Improving Conditional Generation of Fine-Tuned Diffusion Models},
  author={Phunyaphibarn, Prin and Lee, Phillip Y and Kim, Jaihoon and Sung, Minhyuk},
  journal={arXiv preprint arXiv:2503.20240},
  year={2025}
}

@inproceedings{xie2023difffit,
  title={Difffit: Unlocking transferability of large diffusion models via simple parameter-efficient fine-tuning},
  author={Xie, Enze and Yao, Lewei and Shi, Han and Liu, Zhili and Zhou, Daquan and Liu, Zhaoqiang and Li, Jiawei and Li, Zhenguo},
  booktitle={Proceedings of the IEEE/CVF International Conference on Computer Vision},
  pages={4230--4239},
  year={2023}
}

@article{hu2022lora,
  title={Lora: Low-rank adaptation of large language models.},
  author={Hu, Edward J and Shen, Yelong and Wallis, Phillip and Allen-Zhu, Zeyuan and Li, Yuanzhi and Wang, Shean and Wang, Lu and Chen, Weizhu and others},
  journal={ICLR},
  volume={1},
  number={2},
  pages={3},
  year={2022}
}

@inproceedings{moon2022fine,
  title={Fine-tuning diffusion models with limited data},
  author={Moon, Taehong and Choi, Moonseok and Lee, Gayoung and Ha, Jung-Woo and Lee, Juho},
  booktitle={NeurIPS 2022 Workshop on Score-Based Methods},
  year={2022}
}

@inproceedings{hur2024expanding,
  title={Expanding expressiveness of diffusion models with limited data via self-distillation based fine-tuning},
  author={Hur, Jiwan and Choi, Jaehyun and Han, Gyojin and Lee, Dong-Jae and Kim, Junmo},
  booktitle={Proceedings of the IEEE/CVF Winter Conference on Applications of Computer Vision},
  pages={5028--5037},
  year={2024}
}

@article{zhong2024diffusion,
  title={Diffusion tuning: Transferring diffusion models via chain of forgetting},
  author={Zhong, Jincheng and Guo, Xingzhuo and Dong, Jiaxiang and Long, Mingsheng},
  journal={arXiv preprint arXiv:2406.00773},
  year={2024}
}

@article{jensen2025efficient,
  title={Efficient Distillation of Classifier-Free Guidance using Adapters},
  author={Jensen, Cristian Perez and Sadat, Seyedmorteza},
  journal={arXiv preprint arXiv:2503.07274},
  year={2025}
}

@inproceedings{hsiao2024plug,
  title={Plug-and-play diffusion distillation},
  author={Hsiao, Yi-Ting and Khodadadeh, Siavash and Duarte, Kevin and Lin, Wei-An and Qu, Hui and Kwon, Mingi and Kalarot, Ratheesh},
  booktitle={Proceedings of the IEEE/CVF Conference on Computer Vision and Pattern Recognition},
  pages={13743--13752},
  year={2024}
}

@article{zhou2025dice,
  title={DICE: Distilling Classifier-Free Guidance into Text Embeddings},
  author={Zhou, Zhenyu and Chen, Defang and Wang, Can and Chen, Chun and Lyu, Siwei},
  journal={arXiv preprint arXiv:2502.03726},
  year={2025}
}

@article{zhu2022few,
  title={Few-shot image generation with diffusion models},
  author={Zhu, Jingyuan and Ma, Huimin and Chen, Jiansheng and Yuan, Jian},
  journal={arXiv preprint arXiv:2211.03264},
  year={2022}
}

@inproceedings{cao2024few,
  title={Few-shot image generation by conditional relaxing diffusion inversion},
  author={Cao, Yu and Gong, Shaogang},
  booktitle={European Conference on Computer Vision},
  pages={20--37},
  year={2024},
  organization={Springer}
}

@inproceedings{bossard2014food,
  title={Food-101--mining discriminative components with random forests},
  author={Bossard, Lukas and Guillaumin, Matthieu and Van Gool, Luc},
  booktitle={Computer vision--ECCV 2014: 13th European conference, zurich, Switzerland, September 6-12, 2014, proceedings, part VI 13},
  pages={446--461},
  year={2014},
  organization={Springer}
}

@techreport{griffin2007caltech,
  title={Caltech-256 object category dataset},
  author={Griffin, Gregory and Holub, Alex and Perona, Pietro and others},
  year={2007},
  institution={Technical Report 7694, California Institute of Technology Pasadena}
}

@techreport{wah2011caltech,
  title        = {The Caltech-UCSD Birds-200-2011 Dataset},
  author       = {Wah, Catherine and Branson, Steve and Welinder, Peter and Perona, Pietro and Belongie, Serge},
  institution  = {California Institute of Technology},
  number       = {CNS-TR-2011-001},
  year         = {2011},
  type         = {Technical Report},
  address      = {Pasadena, CA, USA}
}

@article{liao2022artbench,
  title={The artbench dataset: Benchmarking generative models with artworks},
  author={Liao, Peiyuan and Li, Xiuyu and Liu, Xihui and Keutzer, Kurt},
  journal={arXiv preprint arXiv:2206.11404},
  year={2022}
}

@inproceedings{krause20133d,
  title={3d object representations for fine-grained categorization},
  author={Krause, Jonathan and Stark, Michael and Deng, Jia and Fei-Fei, Li},
  booktitle={Proceedings of the IEEE international conference on computer vision workshops},
  pages={554--561},
  year={2013}
}

@inproceedings{karras2019style,
  title={A style-based generator architecture for generative adversarial networks},
  author={Karras, Tero and Laine, Samuli and Aila, Timo},
  booktitle={Proceedings of the IEEE/CVF conference on computer vision and pattern recognition},
  pages={4401--4410},
  year={2019}
}

@article{helber2019eurosat,
  title={Eurosat: A novel dataset and deep learning benchmark for land use and land cover classification},
  author={Helber, Patrick and Bischke, Benjamin and Dengel, Andreas and Borth, Damian},
  journal={IEEE Journal of Selected Topics in Applied Earth Observations and Remote Sensing},
  year={2019},
  publisher={IEEE}
}

@article{sadat2023cads,
  title={CADS: Unleashing the diversity of diffusion models through condition-annealed sampling},
  author={Sadat, Seyedmorteza and Buhmann, Jakob and Bradley, Derek and Hilliges, Otmar and Weber, Romann M},
  journal={arXiv preprint arXiv:2310.17347},
  year={2023}
}

@inproceedings{choi2022perception,
  title={Perception prioritized training of diffusion models},
  author={Choi, Jooyoung and Lee, Jungbeom and Shin, Chaehun and Kim, Sungwon and Kim, Hyunwoo and Yoon, Sungroh},
  booktitle={Proceedings of the IEEE/CVF Conference on Computer Vision and Pattern Recognition},
  pages={11472--11481},
  year={2022}
}

@inproceedings{chen2023score,
  title={Score approximation, estimation and distribution recovery of diffusion models on low-dimensional data},
  author={Chen, Minshuo and Huang, Kaixuan and Zhao, Tuo and Wang, Mengdi},
  booktitle={International Conference on Machine Learning},
  pages={4672--4712},
  year={2023},
  organization={PMLR}
}

@article{zheng2023characteristic,
  title={Characteristic guidance: Non-linear correction for diffusion model at large guidance scale},
  author={Zheng, Candi and Lan, Yuan},
  journal={arXiv preprint arXiv:2312.07586},
  year={2023}
}

@article{deja2022analyzing,
  title={On analyzing generative and denoising capabilities of diffusion-based deep generative models},
  author={Deja, Kamil and Kuzina, Anna and Trzcinski, Tomasz and Tomczak, Jakub},
  journal={Advances in Neural Information Processing Systems},
  volume={35},
  pages={26218--26229},
  year={2022}
}

@article{meng2021sdedit,
  title={Sdedit: Guided image synthesis and editing with stochastic differential equations},
  author={Meng, Chenlin and He, Yutong and Song, Yang and Song, Jiaming and Wu, Jiajun and Zhu, Jun-Yan and Ermon, Stefano},
  journal={arXiv preprint arXiv:2108.01073},
  year={2021}
}

@article{heusel2017gans,
  title={Gans trained by a two time-scale update rule converge to a local nash equilibrium},
  author={Heusel, Martin and Ramsauer, Hubert and Unterthiner, Thomas and Nessler, Bernhard and Hochreiter, Sepp},
  journal={Advances in neural information processing systems},
  volume={30},
  year={2017}
}

@article{stein2023exposing,
  title={Exposing flaws of generative model evaluation metrics and their unfair treatment of diffusion models},
  author={Stein, George and Cresswell, Jesse and Hosseinzadeh, Rasa and Sui, Yi and Ross, Brendan and Villecroze, Valentin and Liu, Zhaoyan and Caterini, Anthony L and Taylor, Eric and Loaiza-Ganem, Gabriel},
  journal={Advances in Neural Information Processing Systems},
  volume={36},
  pages={3732--3784},
  year={2023}
}

@article{kynkaanniemi2019improved,
  title={Improved precision and recall metric for assessing generative models},
  author={Kynk{\"a}{\"a}nniemi, Tuomas and Karras, Tero and Laine, Samuli and Lehtinen, Jaakko and Aila, Timo},
  journal={Advances in neural information processing systems},
  volume={32},
  year={2019}
}

@inproceedings{zheng2023improved,
  title={Improved techniques for maximum likelihood estimation for diffusion odes},
  author={Zheng, Kaiwen and Lu, Cheng and Chen, Jianfei and Zhu, Jun},
  booktitle={International Conference on Machine Learning},
  pages={42363--42389},
  year={2023},
  organization={PMLR}
}

@inproceedings{yin2024one,
  title={One-step diffusion with distribution matching distillation},
  author={Yin, Tianwei and Gharbi, Micha{\"e}l and Zhang, Richard and Shechtman, Eli and Durand, Fredo and Freeman, William T and Park, Taesung},
  booktitle={Proceedings of the IEEE/CVF conference on computer vision and pattern recognition},
  pages={6613--6623},
  year={2024}
}

@article{lu2022dpm,
  title={Dpm-solver: A fast ode solver for diffusion probabilistic model sampling in around 10 steps},
  author={Lu, Cheng and Zhou, Yuhao and Bao, Fan and Chen, Jianfei and Li, Chongxuan and Zhu, Jun},
  journal={Advances in Neural Information Processing Systems},
  volume={35},
  pages={5775--5787},
  year={2022}
}

@article{salimans2022progressive,
  title={Progressive distillation for fast sampling of diffusion models},
  author={Salimans, Tim and Ho, Jonathan},
  journal={arXiv preprint arXiv:2202.00512},
  year={2022}
}

@inproceedings{peebles2023scalable,
  title={Scalable diffusion models with transformers},
  author={Peebles, William and Xie, Saining},
  booktitle={Proceedings of the IEEE/CVF international conference on computer vision},
  pages={4195--4205},
  year={2023}
}

@inproceedings{ma2024sit,
  title={Sit: Exploring flow and diffusion-based generative models with scalable interpolant transformers},
  author={Ma, Nanye and Goldstein, Mark and Albergo, Michael S and Boffi, Nicholas M and Vanden-Eijnden, Eric and Xie, Saining},
  booktitle={European Conference on Computer Vision},
  pages={23--40},
  year={2024},
  organization={Springer}
}

@inproceedings{ruiz2023dreambooth,
  title={Dreambooth: Fine tuning text-to-image diffusion models for subject-driven generation},
  author={Ruiz, Nataniel and Li, Yuanzhen and Jampani, Varun and Pritch, Yael and Rubinstein, Michael and Aberman, Kfir},
  booktitle={Proceedings of the IEEE/CVF conference on computer vision and pattern recognition},
  pages={22500--22510},
  year={2023}
}

@article{shen2025efficient,
  title={Efficient Diffusion Models: A Survey},
  author={Shen, Hui and Zhang, Jingxuan and Xiong, Boning and Hu, Rui and Chen, Shoufa and Wan, Zhongwei and Wang, Xin and Zhang, Yu and Gong, Zixuan and Bao, Guangyin and others},
  journal={arXiv preprint arXiv:2502.06805},
  year={2025}
}

@article{weiss2016survey,
  title={A survey of transfer learning},
  author={Weiss, Karl and Khoshgoftaar, Taghi M and Wang, DingDing},
  journal={Journal of Big data},
  volume={3},
  pages={1--40},
  year={2016},
  publisher={Springer}
}

\clearpage
\appendix
\section*{Appendix}

%
%
%
%

\section{Additional Methodology Explanations}

\subsection{DogFit Algorithm}
\label{app:algorithm}

Algorithm~\ref{alg:DogFit} summarizes the training loop of \texttt{DogFit} with controllability through guidance conditioning and the scheduling strategies.

\begin{algorithm}[ht]
\caption{\texttt{DogFit} training with Controllable Guidance and Scheduling}
\label{alg:DogFit}
\begin{algorithmic}[1]
\REQUIRE Target dataset $\{\mathcal{D^T}\}$, noise schedule $\bar{\alpha}$, fixed source model $\epsilon_{\theta_0}$, training model $\epsilon_\theta$ initialized by $\epsilon_{\theta_0}$, guidance cut-off $\tau_{\text{c}}$, late-start step $\tau_{\text{s}}$. 
\FOR{training step $s = 1$ to $S$}
    \STATE Sample target data $(x_0, c) \sim \{\mathcal{D^T}\}$
    \STATE Sample noise $\epsilon \sim \mathcal{N}(0, 1)$ and time-step $t \sim \mathcal{U}(0, 1)$
    \STATE Sample guidance strength $\boldsymbol{w}=1+z, \quad z \sim \mathcal{P}(z)$
    \STATE Add noise: $x_t = \sqrt{\bar{\alpha}_t} x_0 + \sqrt{1 - \bar{\alpha}_t} \cdot \epsilon$
    \IF{$s > \tau_{\text{s}}$ \AND $t < \tau_{\text{c}}$}
        \STATE Compute guided target: \\
        $\epsilon' = \epsilon + (\boldsymbol{w} - 1) \cdot \text{sg}\left( \epsilon_\theta(x_t \mid c, \boldsymbol{1}, \mathcal{D^T}) - \epsilon_{\theta_0}(x_t) \right)$
    \ELSE
        \STATE $\epsilon' = \epsilon$
    \ENDIF
    \STATE Compute loss: 
    $\mathcal{L}_{\text{\texttt{DogFit}}} = \left\| \epsilon_\theta(x_t \mid c, \boldsymbol{w}, \mathcal{D^T}) - \epsilon' \right\|^2$
    \STATE Backpropagate and update: $\theta = \theta - \eta \nabla_\theta \mathcal{L}_{\text{\texttt{DogFit}}}$
\ENDFOR
\end{algorithmic}
\end{algorithm}

\subsection{CDF of Sampled Guidance Strengths}
\label{app:cdf}

\begin{figure}[t]
    \centering
    \includegraphics[width=\linewidth]{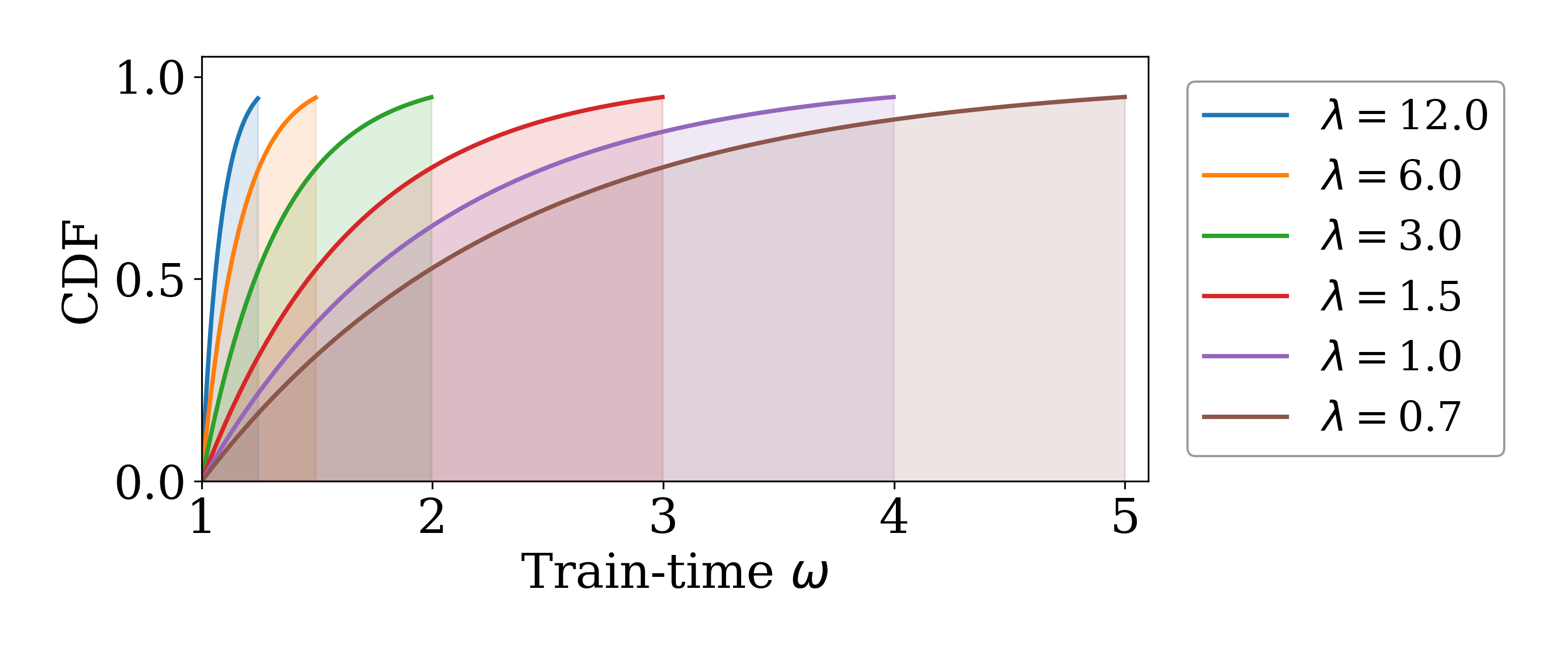}
    \caption{The CDF of sampled guidance strengths \( w \) during training for various \( \lambda \) values using SEDD. The shaded region covers 95\% of the samples.}
    \label{fig:cdf_}
\end{figure}

Fig.~\ref{fig:cdf_} visualizes the Cumulative Distribution Function (CDF) of \( w \)s sampled during training from the shifted exponentially decaying distribution (SEDD) in Equation~\ref{eq:exponential}. Larger values of \( \lambda \) concentrate the sampling mass closer to \( w = 1 \), limiting exposure to higher guidance values during training. This design ensures that the model primarily learns under weak to moderate guidance, while still sparsely encountering stronger guidance values.

\subsection{Theoretical Insights into DogFit}
\label{app:theory}
In this section, we motivate the shift from using the marginal target noise estimate to the marginal source estimate in transfer learning. We discuss the benefits of this shift, and justify our method of applying it in the fine-tuning process. Then we show that under mild assumptions, \texttt{DogFit} learns to internalize DoG's behaviour. Further, we conjecture that \texttt{DogFit} is an extension of MG, with an added domain alignment term. 

\vspace{0.5em}
\noindent\textbf{General Assumptions.} We assume that both the pretrained source model $\epsilon_{\theta_0}$ and the fine-tuned model $\epsilon_{\theta}$ share the same architecture and noise schedule $\alpha_t$, and are trained to approximate the same family of score functions under a common sampling procedure.

\paragraph{Guidance via Unconditional Source Prior}
The guidance signal in CFG is originally created via the target domain conditional posterior $p_\theta(c \mid x_t, \mathcal{D}^{\mathcal{T}})$ to encourage sample fidelity and strong class-conditional alignment in the target domain. By Bayes' rule:
\begin{equation}
p_\theta(c \mid x_t, \mathcal{D}^{\mathcal{T}}) = \frac{p_\theta(x_t \mid c, \mathcal{D}^{\mathcal{T}}) \cdot p_\theta(c \mid \mathcal{D}^{\mathcal{T}})}{p_\theta(x_t \mid \mathcal{D}^{\mathcal{T}})},
\end{equation}


\noindent or equivalently, in the log form:
\begin{align}
\log &p_\theta(c \mid x_t, \mathcal{D}^{\mathcal{T}})=
\\&  \log p_\theta(x_t \mid c, \mathcal{D}^{\mathcal{T}})
 - \log p_\theta(x_t \mid \mathcal{D}^{\mathcal{T}}) + \log p_\theta(c \mid \mathcal{D}^{\mathcal{T}}). \nonumber
\end{align}

We consider the score functions (gradients of log-densities):
\begin{align}
\nabla_{x_t} \log p_\theta&(c \mid x_t, \mathcal{D}^{\mathcal{T}})
=\\&  \nabla_{x_t} \log p_\theta(x_t \mid c, \mathcal{D}^{\mathcal{T}})
 - \nabla_{x_t} \log p_\theta(x_t \mid \mathcal{D}^{\mathcal{T}}), \nonumber
\end{align}
which are approximated by diffusion models:
\begin{align}
\label{equation:noise_finetuned}
\nabla_{x_t} \log p_\theta(x_t \mid c, \mathcal{D}^{\mathcal{T}}) & \approx -\frac{1}{\sigma_t} \epsilon_\theta(x_t \mid c, \mathcal{D}^{\mathcal{T}}), \\
\label{equation:noise_pretrained}
\nabla_{x_t} \log p_\theta(x_t \mid \mathcal{D}^{\mathcal{T}}) & \approx -\frac{1}{\sigma_t} \epsilon_\theta(x_t \mid \mathcal{D}^{\mathcal{T}}).
\end{align}

Traditionally, CFG estimates both the class-conditional (Eq. \ref{equation:noise_finetuned}) and the marginal (Eq. \ref{equation:noise_pretrained}) noises with the same diffusion model. With enough data in the target domain, the class-conditional noise can be well estimated during fine-tuning via $\epsilon_\theta(x_t \mid c, \mathcal{D}^{\mathcal{T}})$. However, learning the target noise estimator $\epsilon_\theta(x_t \mid \mathcal{D}^{\mathcal{T}})$ faces the risk of under-training due to only being trained on a small portion of the tiny target dataset (typically 10\% ~\cite{ho2022classifier}. 
Since the source model is usually trained on large, diverse datasets, it likely captures a good approximation of the general image manifold. Furthermore, it has been shown that diffusion models have generalizability on other data distributions \cite{deja2022analyzing} and can denoise noisy images from domains other than those they have been trained on \cite{meng2021sdedit}. Therefore, we can assume that the pre-trained source model captures a general enough prior,
giving us the ability to approximate the unconditional score using the pre-trained source model. This leads us to the formulation of DoG:
\begin{equation}
\nabla_{x_t} \log p_{\theta}(x_t \mid \mathcal{D}^{\mathcal{T}}) \approx -\frac{1}{\sigma_t} \epsilon_{\theta_0}(x_t).
\end{equation}

Thus, the posterior score can be approximated as:
\begin{align}
\nabla_{x_t} \log p_\theta(c \mid x_t, \mathcal{D}^{\mathcal{T}})
&\propto -\frac{1}{\sigma_t} \left( \epsilon_\theta(x_t \mid c, \mathcal{D}^{\mathcal{T}}) - \epsilon_{\theta_0}(x_t) \right),
\end{align}
\noindent which exactly matches the offset used in DoG sampling~\citep{zhong2025domain}. Our method takes a step further from this and applies the same guidance offset online. We train a diffusion model to directly predict this guided score instead of separately learning Eq. \ref{equation:noise_finetuned} and Eq. \ref{equation:noise_pretrained} in the form of DoG.

\paragraph{Proposition 1.} Let $\epsilon_\theta(x_t \mid c, w)$ be the \texttt{DogFit} model with controllable guidance trained with the objective:
\begin{align}
\mathcal{L}&_{\text{\texttt{DogFit}}} = \\& \left\| \epsilon_\theta(x_t \mid c, w, \mathcal{D}^{\mathcal{T}}) - \left( \epsilon + (w - 1) \cdot \Delta(x_t, c,\mathcal{D}^{\mathcal{T}}) \right) \right\|^2,\nonumber
\end{align}
where $\Delta(x_t, c, \mathcal{D}^{\mathcal{T}}) = \epsilon_\theta(x_t \mid c,w=1,\mathcal{D}^{\mathcal{T}}) - \epsilon_{\theta_0}(x_t)$ and assume the following conditions hold after convergence:
\begin{enumerate}
    \item[(i)] When no guidance is applied, the model approximately recovers the true noise $\epsilon$ used in training.
    
    \item[(ii)] The model's prediction varies linearly with the guidance strength $w$.
\end{enumerate}

Then for any $w$, the \texttt{DogFit} model matches the DoG-guided score used at sampling time:
\begin{align}
& \epsilon_\theta(x_t \mid c, w, \mathcal{D}^{\mathcal{T}}) = \\& \epsilon_\theta(x_t \mid c, \mathcal{D}^{\mathcal{T}}) + (w - 1) \cdot \left( \epsilon_\theta(x_t \mid c, \mathcal{D}^{\mathcal{T}}) - \epsilon_{\theta_0}(x_t) \right). \nonumber
\end{align}

\noindent\textit{- Note on the assumptions:} Prior work on CFG has shown that the linearity assumption holds in practice for low to moderate $w$ values~\cite{zheng2023characteristic}. Furthermore, our use of an SEDD to sample $w$ during training (see Sec. 3~\ref{sec:controllable_guidance}) helps maintain accurate estimation of $\epsilon$ in the unconditional setting. Based on this, we conjecture that both assumptions approximately hold true in practice.

\vspace{0.5em}
\noindent\textbf{Proof sketch.} \quad
At convergence, the \texttt{DogFit} objective minimizes the expected loss:
\begin{align}
&\mathbb{E}_{x_0, t, \epsilon, w} \\& \left[ \left\| \epsilon_\theta(x_t \mid c, w, \mathcal{D}^{\mathcal{T}}) - \left( \epsilon + (w - 1) \cdot \Delta(x_t, c, \mathcal{D}^{\mathcal{T}}) \right) \right\|^2 \right]. \nonumber
\end{align}

Assuming that $\epsilon_\theta(x_t \mid c, w)$ is linear in $w$, we can write the model as:
\begin{align}
\epsilon_\theta(x_t \mid c, w, \mathcal{D}^{\mathcal{T}}) = A + B \cdot (w - 1).
\end{align}

Since the training loss is a squared error between two linear functions of $w$, minimizing this loss for all values of $w$ is equivalent to matching the corresponding coefficients of both functions. This means that to minimize the loss over the full support of $w$, the model must satisfy:
\begin{align}
A = \epsilon, \quad B = \Delta(x_t, c, \mathcal{D}^{\mathcal{T}}).
\end{align}

Since $\epsilon_\theta(x_t \mid c, \mathcal{D}^{\mathcal{T}}) \approx \epsilon_\theta(x_t \mid c,w=1, \mathcal{D}^{\mathcal{T}}) \approx \epsilon$, this gives the optimal solution:
\begin{align}
\epsilon_\theta(x_t \mid c, w, \mathcal{D}^{\mathcal{T}}&) =
\\&\epsilon + (w - 1) \cdot \left( \epsilon_\theta(x_t \mid c, \mathcal{D}^{\mathcal{T}}) - \epsilon_{\theta_0}(x_t) \right), \nonumber
\end{align}
which matches the DoG direction used at test time. Therefore, \texttt{DogFit} learns to internalize the DoG guidance signal during training, removing the need for external guidance at sampling. 
\hfill $\blacksquare$

\paragraph{Proposition 2.} \texttt{DogFit} implicitly augments MG's training process with a classifier-based domain alignment term:
\begin{equation}
\label{eq:domain_alignment}
    \epsilon'_\text{\texttt{DogFit}} = \epsilon'_\text{MG} - \sigma (w-1) \cdot \nabla_{x_t} \log p_{\theta}(\mathcal{D^T}|x_t).
\end{equation}

This enables \texttt{DogFit} to retain MG’s class-conditional guidance while avoiding out-of-distribution generation by pulling samples the target domain manifold's core.

\paragraph{Proof sketch.} The target noise in MG, when applied in the task of fine-tuning, can be rewritten as:
\begin{equation}
    \epsilon'_\text{MG} = \epsilon + (w-1)\cdot \text{sg} \left( \epsilon_{\theta}(x_t|c, \mathcal{D^T}) - \epsilon_{\theta}(x_t|
    \mathcal{D^T}) \right).
\end{equation}

Then, by subtracting the target noise values of MG and \texttt{DogFit}, we have:
\begin{equation}
\label{equation:proof1}
    \epsilon'_\text{\texttt{DogFit}} - \epsilon'_\text{MG} = (w-1)\cdot \text{sg} \left( \epsilon_{\theta}(x_t|\mathcal{D^T}) - \epsilon_{\theta_0}(x_t) \right).
\end{equation}

According to Eq. \ref{equation:noise_finetuned} and Eq. \ref{equation:noise_pretrained}, Eq. \ref{equation:proof1} can be rewritten:
\begin{align}
\epsilon&'_\text{\texttt{DogFit}} - \epsilon'_\text{MG} =
    \\& -\sigma (w-1)\cdot \text{sg} \left(  \nabla_{x_t} \log p_{\theta}(x_t | \mathcal{D^T}) - \nabla_{x_t} \log p_{\theta}(x_t) \right). \nonumber
\end{align}

By Bayes' rule, we have:
\begin{equation}
\frac{p(x_t|\mathcal{D^T})}{p(x_t)} \propto p(\mathcal{D^T}|x_t),
\end{equation}
which is equivalent to:
\begin{equation}
     \nabla_{x_t} \log p(\mathcal{D^T}|x_t) = \nabla_{x_t} \log p_t(x_t | \mathcal{D^T}) - \nabla_{x_t} \log p_t(x_t).
\end{equation}

 Thus, given that MG uses $\epsilon_{\theta}(x_t|\mathcal{D^T})$ as an approximation of the term $-\sigma_t \nabla_{x_t} \log p(x_t| \mathcal{D^T})$, we can express \texttt{DogFit} as applying the MG's direction, but with a correction term that further pushes the sample towards a target domain classifier's gradient direction.
 \hfill $\blacksquare$

\begin{figure*}[t]
    \centering
        \includegraphics[width=\linewidth]{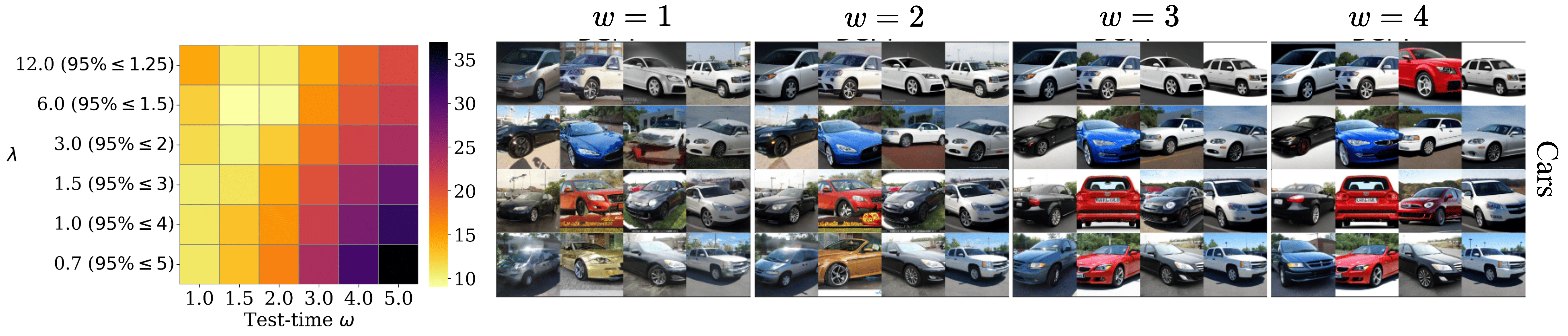} \\
                \caption{Effect of controllable guidance at test time for the Stanford-Cars domain. (Left) FID heatmaps showing the behaviour of FID given different test-time $w$s with different training-time $\lambda$s. (Right) Corresponding generated samples for fixed $\lambda = 3$ (i.e., 95\% of sampled $w$ values lie between 1 and 2), across varying test-time $w$.}
                \label{fig:controllable_guidane_plot_cars}
\end{figure*}

\begin{figure*}[t]
    \centering

    \begin{minipage}[t]{0.49\textwidth}
        \centering
        \includegraphics[width=\linewidth]{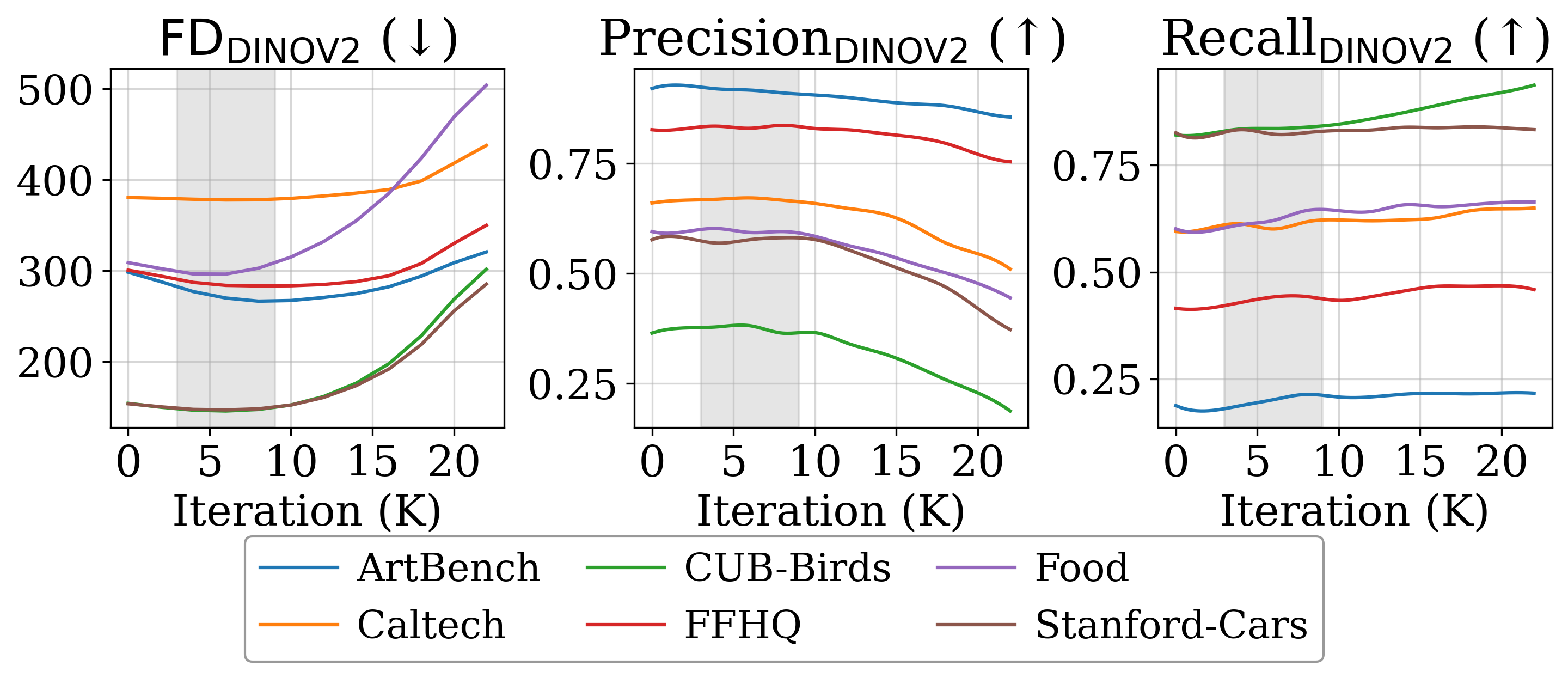}
        \vspace{0.5em}
        \textbf{(a)} Late-start ablation
    \end{minipage}
    \hfill
    \begin{minipage}[t]{0.49\textwidth}
        \centering
        \includegraphics[width=\linewidth]{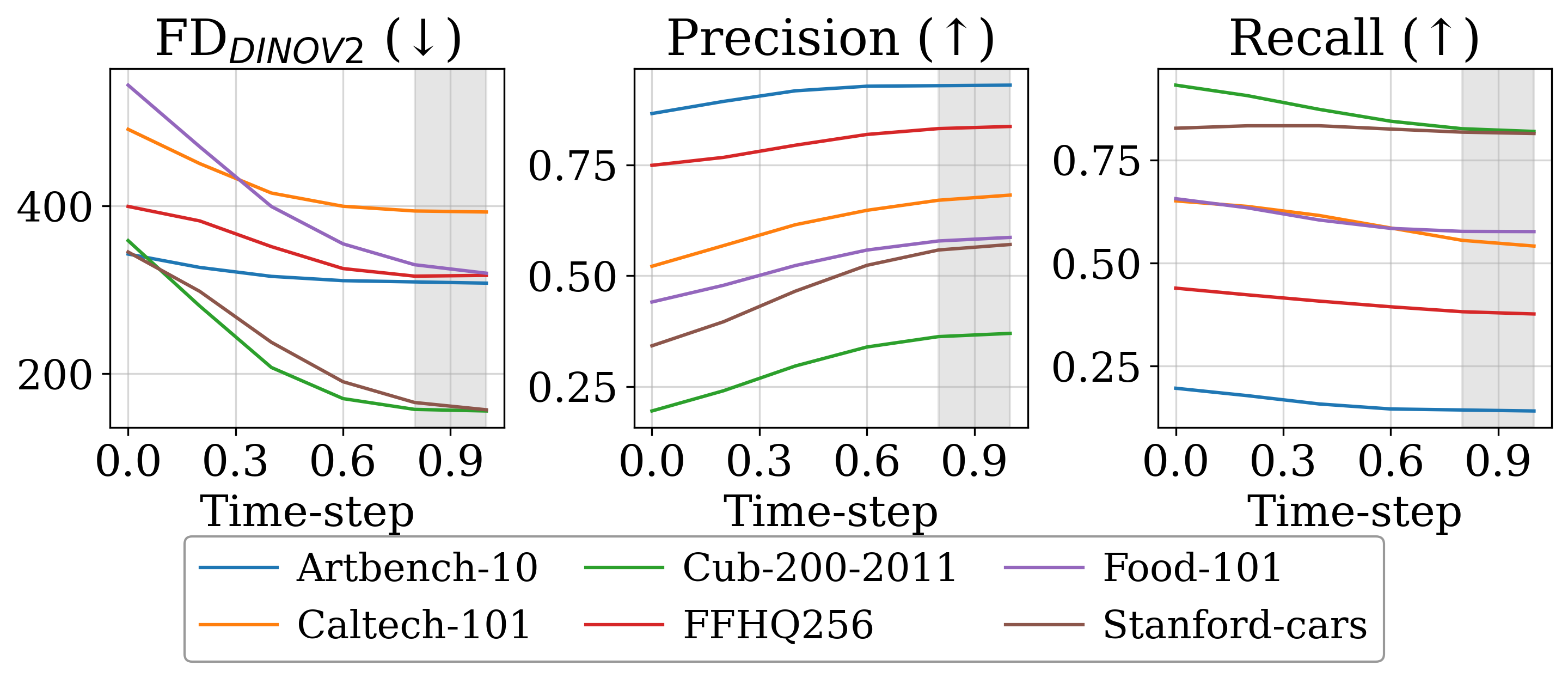}
        \vspace{0.5em}
        \textbf{(b)} Cut-off ablation
    \end{minipage}

    \caption{Ablation on guidance schedules in \texttt{DogFit}. (a) Varying the late-start threshold $\tau_{\text{s}}$ to control when guidance begins. (b) Varying the cut-off threshold $\tau_{\text{c}}$ to restrict guidance to later denoising steps. Performed using FD$_{\text{DINOV2}}$ on DiT-XL/2.}
    \label{fig:ablation_grid_dino}
\end{figure*}

\begin{figure}[t]
    \centering
    \includegraphics[width=\linewidth]{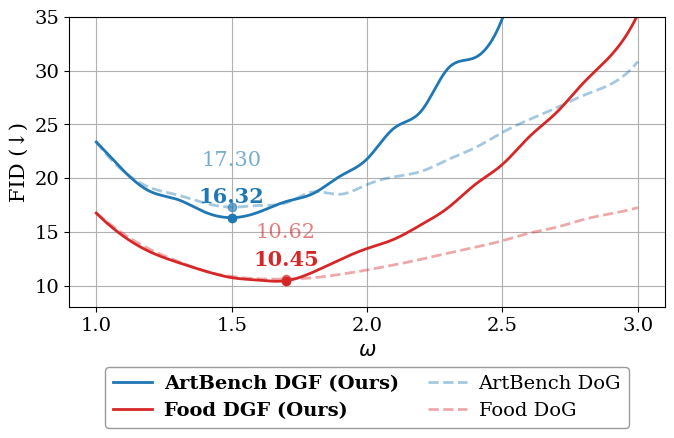}
    \caption{Comparing \texttt{DogFit} (without guidance control) and DoG in sensitivity of FID to $w$. }
    \label{fig:ablation_guidance_strength}
\end{figure}

\begin{figure}[t]
    \centering
    \includegraphics[width=\linewidth]{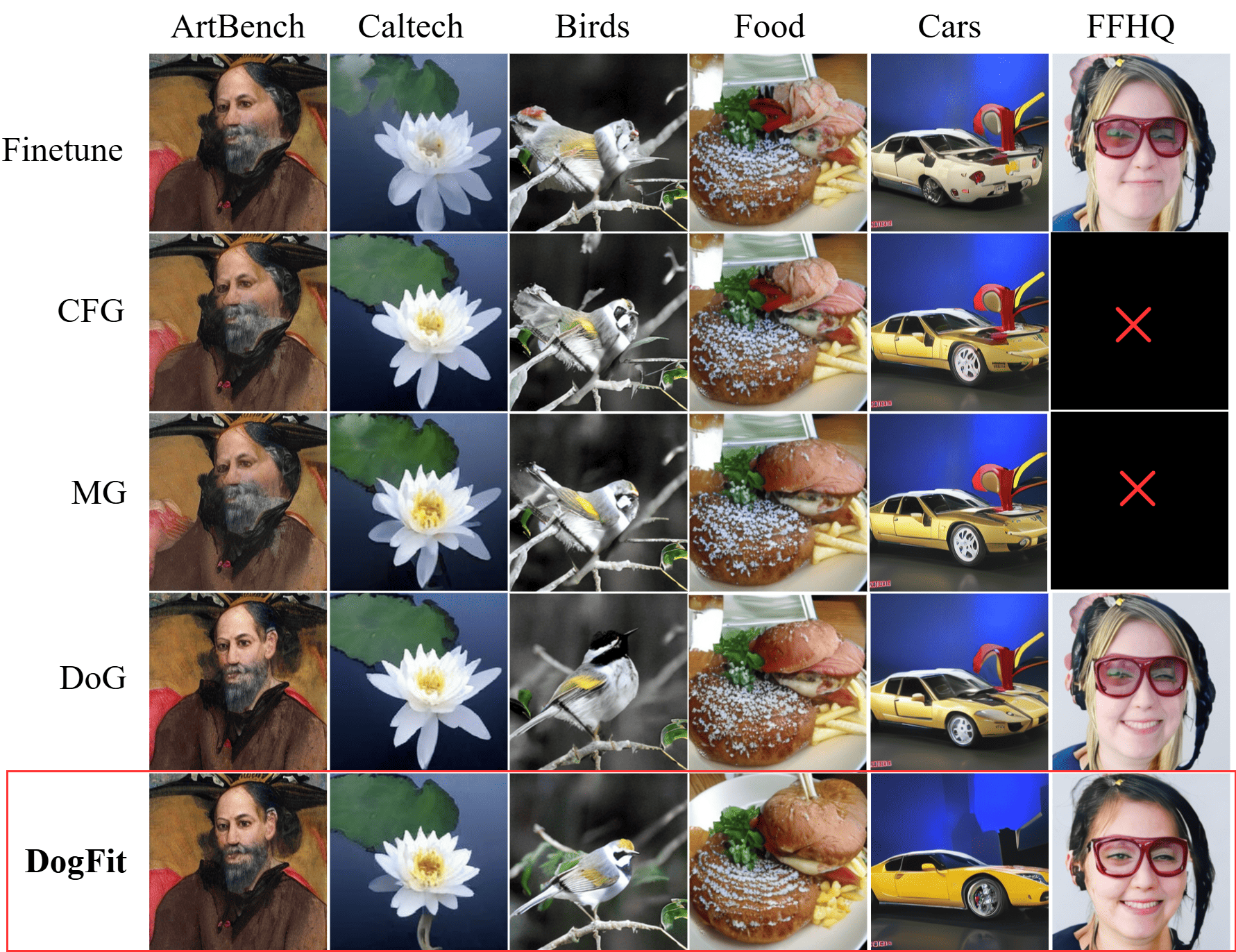}
    \caption{Qualitative comparison of guidance mechanisms on SiT-XL/2 (guidance scale 1.5). $\textcolor{red}{\times}$: not applicable.}
    \label{fig:qualitative_comparison2}
\end{figure}

\section{Additional Experiments}
\label{app:experiments}

\subsection{FID Sensitivity to $w$ without Controllability} \label{app:exp_sensitivity}
We further examine the sensitivity of our method to $w$ in the absence of controllability. We train a specific model for each $w$ and compare the FID trends of \texttt{DogFit} and DoG across varying values (Fig.~\ref{fig:ablation_guidance_strength}). While both methods achieve optimal FID in a similar range, \texttt{DogFit} exhibits a sharper degradation for larger $w$, suggesting higher sensitivity to strong guidance values. We hypothesize that this is due to training-time overexposure to extreme guidance, which can lead the model to learn from unnatural signals that do not follow the real target distribution, thereby reducing generation quality. This motivates our choice of sampling $w$ from an exponentially decaying distribution, ensuring the model is only sparsely exposed to high guidance values during training.

\subsection{Controllable Guidance Analysis on Cars} \label{app:exp_control} 
Fig.~\ref{fig:controllable_guidane_plot_cars} provides an extended version of the controllable guidance analysis for the Stanford-Cars dataset. This complements the main findings and illustrates the same trends on a new domain.

\begin{table*}[ht]
\centering
\caption{Precision and Recall results on DiT/XL-2 backbone using Inception and DINOV2 metrics.}
\label{tab:Precision_Recall_combined_vertical}
\renewcommand{\arraystretch}{1.3}
\resizebox{\textwidth}{!}{
\begin{tabular}{|c|l|c|ccccc|c|}
\hline
\multirow{2}{*}{\textbf{Metric}} & \multirow{2}{*}{\textbf{Method}} & \textbf{Unlabelled} & \multicolumn{6}{c|}{\textbf{Labelled}} \\
\cline{3-9}
 &  & FFHQ & ArtBench & Caltech & CUB-Birds & Food & Stanford-Cars & Avg. \\
\hline
\multirow{6}{*}{Precision (Inception)} 
& Fine-tuning & 0.70 & 0.67 & 0.60 & 0.58 & 0.79 & 0.51 & 0.63 \\
& + CFG \cite{ho2022classifier} & - & 0.68 & 0.76 & {0.72} & 0.82 & 0.63 & 0.72 \\
& + DoG \cite{zhong2025domain} & \textbf{0.83} & \textbf{0.77} & \textbf{0.81} & \underline{0.82} & \textbf{0.88} & \underline{0.77} & \textbf{0.81} \\
& MG \cite{tang2025diffusion} & - & 0.68 & 0.76 & 0.71 & 0.81 & 0.62 & 0.72 \\
 & \texttt{DogFit} \cellcolor{gray!10}  & \underline{0.81} & \underline{0.76} & 0.75 & \underline{0.82} & \underline{0.87} & {0.74} & \underline{0.79} \\
 & \texttt{DogFit + Control} \cellcolor{gray!10}  & \textbf{0.83} & \underline{0.76} & \underline{0.78} & \textbf{0.83} & \underline{0.87} & \textbf{0.80} & \textbf{0.81} \\
\hline
\multirow{6}{*}{Precision (DINOV2)} 
& Fine-tuning & 0.74 & 0.84 & 0.48 & 0.16 & 0.41 & 0.30 & 0.44 \\
& + CFG \cite{ho2022classifier} & - & 0.83 & \underline{0.67} & 0.30 & 0.48 & 0.47 & 0.52 \\
& + DoG \cite{zhong2025domain} & {0.82} & \underline{0.91} & \textbf{0.73} & \textbf{0.39} & \textbf{0.59} & \textbf{0.61} & \textbf{0.65} \\
& MG \cite{tang2025diffusion} & - & 0.84 & 0.63 & 0.28 & 0.51 & 0.44 & 0.54 \\
& \texttt{DogFit} \cellcolor{gray!10}  & \underline{0.83} & \textbf{0.92} & \underline{0.67} & \underline{0.38} & \underline{0.58} & {0.58} & \underline{0.62} \\
& \texttt{DogFit + Control} \cellcolor{gray!10}  & \textbf{0.86} & {0.90} & \underline{0.67} & \underline{0.38} & \underline{0.58} & \underline{0.59} & \underline{0.62} \\
\hline \hline
\multirow{6}{*}{Recall (Inception)} 
& Fine-tuning & \textbf{0.67} & \textbf{0.57} & \textbf{0.75} & \textbf{0.81} & 0.52 & \underline{0.58} & \textbf{0.65} \\
& + CFG \cite{ho2022classifier} & - & \underline{0.56} & 0.66 & \underline{0.76} & \underline{0.53} & \textbf{0.59} & 0.62 \\
& + DoG \cite{zhong2025domain} & 0.61 & 0.50 & 0.65 & 0.68 & 0.48 & 0.52 & 0.57 \\
& MG \cite{tang2025diffusion} & - & \underline{0.56} & 0.67 & 0.75 & \textbf{0.55} & \textbf{0.59} & \underline{0.624} \\
 & \texttt{DogFit} \cellcolor{gray!10} & \underline{0.63} & 0.51 & \underline{0.70} & 0.68 & 0.50 & 0.55 & 0.59 \\
 & \texttt{DogFit + Control} \cellcolor{gray!10} & {0.57} & {0.51} & {0.68} & {0.67} & {0.50} & {0.48} & {0.57} \\
\hline
\multirow{6}{*}{Recall (DINOV2)} 
& Fine-tuning & \textbf{0.46} & 0.21 & \textbf{0.66} & \textbf{0.95} & \textbf{0.67} & 0.82 & \textbf{0.66} \\
& + CFG \cite{ho2022classifier} & - & \textbf{0.22} & 0.59 & 0.87 & 0.61 & \textbf{0.84} & \underline{0.63} \\
& + DoG \cite{zhong2025domain} & \underline{0.44} & 0.21 & 0.57 & 0.83 & \underline{0.64} & 0.83 & 0.62 \\
& MG \cite{tang2025diffusion} & - & \textbf{0.22} & {0.60} & \textbf{0.89} & 0.60 & \textbf{0.84} & \underline{0.63} \\
 & \texttt{DogFit} \cellcolor{gray!10} & 0.43 & 0.21 & {0.60} & 0.83 & \underline{0.64} & 0.83 & 0.62 \\
 & \texttt{DogFit + Control} \cellcolor{gray!10} & {0.43} & {0.21} & \underline{0.61} & {0.85} & \underline{0.64} & {0.83} & \underline{0.63} \\
\hline
\end{tabular}
}
\end{table*}

\begin{table}[t] \centering \scriptsize \begin{tabular}{|c|c c c c | c|} \hline \textbf{Test-time} & \multicolumn{4}{c|}{\textbf{Uniform}} & \multicolumn{1}{c|}{\cellcolor{gray!10}\textbf{SEDD}}\\ \cline{2-5} \textbf{$\mathbf{\omega}$} & $\mathcal{U}[1,1.1]$ & $\mathcal{U}[1,1.25]$ & $\mathcal{U}[1,1.5]$ & $\mathcal{U}[1,2]$ & \textbf{\cellcolor{gray!10}($\mathbf{\lambda=2}$)} \\ \hline\hline
1 & 19.02 & 16.68 & 20.08 & 22.08 & 12.98 \\
\cellcolor{gray!10}1.5 & 19.62 & 13.84 & 18.27 & 18.05 & \textbf{10.94} \\
2 & 29.65 & 15.97 & 26.29 & 24.23 & 13.05 \\
3 & 61.56 & 28.28 & 53.28 & 48.00 & 19.81 \\
4 & 85.62 & 53.24 & 70.60 & 65.12 & 24.84 \\
5 & 114.97 & 98.75 & 88.39 & 77.53 & 27.95 \\
\hline \end{tabular}
\caption{Ablating training-time guidance sampling methods, by comparing SEDD and uniform sampling. Results indicate FID$\downarrow$ across a wide range of test-time $\omega$ values.} \label{tab:uniform_sed} \end{table}

\subsection{Guidance Scheduling on $\text{FD}_\text{DINOV2}$}
\label{app:fd_dino}
We observe that delaying guidance using the late-start strategy improves both $\text{FD}_\text{DINOV2}$ and precision, consistent with our FID-based findings. This supports the idea that postponing guidance allows for stronger and more accurate directional signals once the model has sufficiently stabilized (Fig.~\ref{fig:ablation_grid_dino}a).
However, unlike FID, applying the cut-off strategy (limiting guidance to later denoising steps) shows little to no improvement in $\text{FD}\text{DINOV2}$ across most domains (Fig.~\ref{fig:ablation_grid_dino} (b)). We believe this is due to a fundamental difference in what the two metrics capture: FID benefits from preserving the global structure inherited from the source model, which is often formed early in the denoising process. In contrast, $\text{FD}\text{DINOV2}$ focuses more on semantic alignment with the target domain, which may require guidance to be present throughout the generation, not just in the later steps. This distinction further highlights the importance of using multiple metrics when evaluating generative models.

\subsection{Qualitative Comparison on SiT-XL/2} 
\label{app:exp_sit}
Fig.~\ref{fig:qualitative_comparison2} provides a qualitative comparison on the SiT-XL/2 backbone. \texttt{DogFit} generates sharp, domain-relevant details on
par with DoG, while avoiding out-of-distribution
behavior occasionally seen with CFG and MG (e.g., distorted faces and birds).

\begin{figure*}[t]
    \centering
        \includegraphics[width=\linewidth]{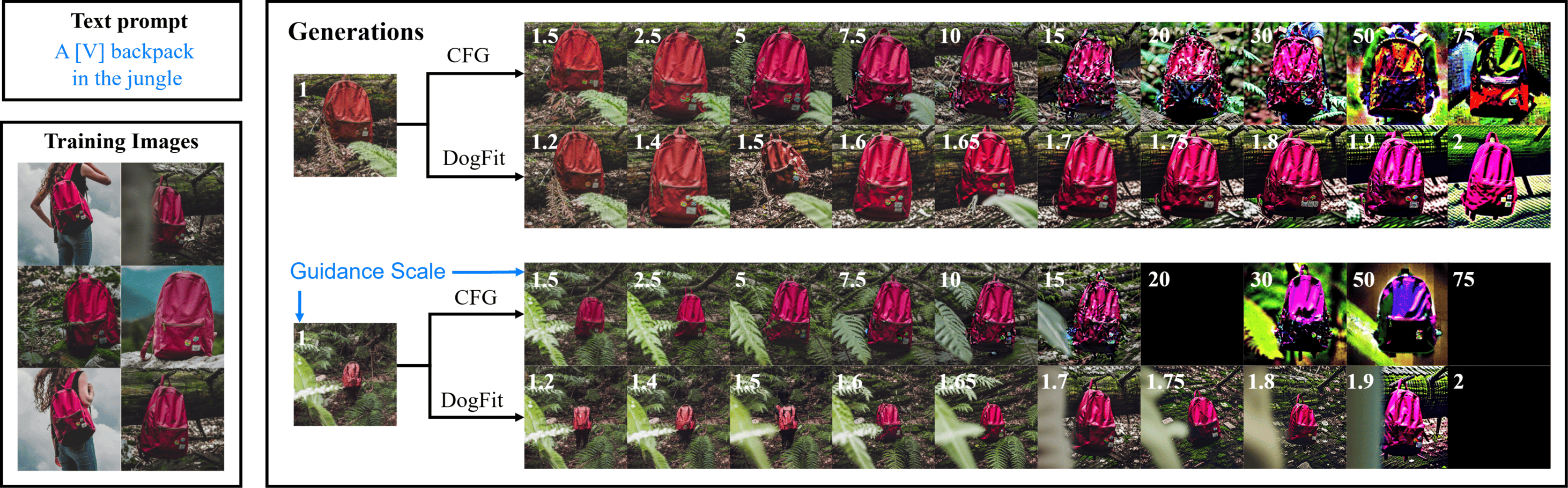} \\
                \caption{Qualitative DreamBooth text-to-image generations (SD~v1.5) comparing CFG and \texttt{DogFit}. Black images correspond to Not-Safe-For-Work (NSFW) content detection using hugging face generators.}
                \label{fig:dreambooth_qualitative}
\end{figure*}

\subsection{Precision and Recall Analysis}
\label{app:exp_precision_recall}
Tab.\ref{tab:Precision_Recall_combined_vertical} presents a comparison of Precision and Recall~\citep{kynkaanniemi2019improved} between \texttt{DogFit} and the baselines on the DiT-XL/2 backbone. 
Precision reflects the fidelity or quality of generated samples, while Recall measures sample diversity. As expected, all guidance methods consistently improve Precision compared to fine-tuning while reducing Recall, aligning with the general behavior of guided diffusion models. This effect arises because guidance steers sampling toward high-likelihood regions of the conditional distribution, which inherently narrows the output distribution and suppresses diversity.
Our method, both with and without guidance control, matches the performance of DoG in terms of both Precision and Recall in most settings. Notably, \texttt{DogFit} achieves the highest or second-highest Precision in both Inception and DINOV2 metrics across all datasets).
In terms of Recall, \texttt{DogFit} and DoG slightly under-perform CFG and MG, which we believe is due to the additional domain alignment term further reducing the diversity to prevent out-of-domain target generations (Eq. \ref{eq:domain_alignment}). While recent works explore diversity-preserving techniques and scheduling mechanisms for guidance during sampling~\citep{kynkaanniemi2024applying, sadat2023cads}, analogous techniques for training-time guidance are yet to be developed. We leave the exploration of such methods for \texttt{DogFit} as an exciting direction for future work.

\begin{table}[t]
\centering
\scriptsize
\resizebox{\linewidth}{!}{%
\begin{tabular}{|l|c c c c|}
\hline
\textbf{Method} & \multicolumn{1}{c}{\textbf{FID↓}} &
\multicolumn{1}{c}{\textbf{KID↓} ($\times 10^{-4}$)} &
\multicolumn{1}{c}{\textbf{$\text{FD}_{\text{DINOv2}}$↓}} &
\multicolumn{1}{c|}{\textbf{\#Pass}} \\
\cline{2-5}
\hline\hline
Fine-tuning & 15.73 & 90.41 & 510.60 & \textbf{x1} \\
+ CFG & 11.93 {\tiny (\textcolor{green!50!black}{+24.16\%})}
      & 57.69 {\tiny (\textcolor{green!50!black}{+36.19\%})}
      & \textbf{366.00} {\tiny (\textcolor{green!50!black}{+28.32\%})}
      & x2 \\
\cellcolor{gray!10} \texttt{DogFit}
      & \textbf{11.06} {\tiny (\textcolor{green!50!black}{+29.69\%})}
      & \textbf{43.15} {\tiny (\textcolor{green!50!black}{+52.27\%})}
      & 380.44 {\tiny (\textcolor{green!50!black}{+25.49\%})}
      & \textbf{x1} \\
\hline
\end{tabular}%
}
\caption{Adapting DiT-XL/2 to Food101 at $512{\times}512$ resolution. Percent improvements relative to Fine-tuning are in parentheses. }
\label{tab:food512}
\end{table}

\begin{table}[t]
\centering
\scriptsize
\resizebox{\linewidth}{!}{%
\begin{tabular}{|l|c c c c|}
\hline
\textbf{Method} & \multicolumn{1}{c}{\textbf{FID↓}} &
\multicolumn{1}{c}{\textbf{KID↓} ($\times 10^{-4}$)} &
\multicolumn{1}{c}{\textbf{$\text{FD}_{\text{DINOv2}}$↓}} &
\multicolumn{1}{c|}{\textbf{\#Pass}} \\
\cline{2-5}
\hline\hline
Fine-tuning & 36.08 & 24.23 & 258.62 & \textbf{x1} \\
+ CFG
  & 33.38 {\tiny (\textcolor{green!50!black}{+7.48\%})}
  & 21.25 {\tiny (\textcolor{green!50!black}{+12.30\%})}
  & 257.73 {\tiny (\textcolor{green!50!black}{+0.34\%})}
  & x2 \\
\cellcolor{gray!10} \texttt{DogFit}
  & \textbf{29.30} {\tiny (\textcolor{green!50!black}{+18.79\%})}
  & \textbf{16.51} {\tiny (\textcolor{green!50!black}{+31.86\%})}
  & \textbf{221.03} {\tiny (\textcolor{green!50!black}{+14.53\%})}
  & \textbf{x1} \\
\hline
\end{tabular}
}
\caption{Adapting DiT-XL/2 to the distant EuroSat domain at $256{\times}256$.}
\label{tab:eurosat}
\end{table}

\begin{table}[t]
\centering
\scriptsize
\resizebox{\linewidth}{!}{%
\begin{tabular}{|l|c c c c|}
\hline
\textbf{Method} &
\textbf{DINO $\uparrow$} &
\textbf{CLIP-I $\uparrow$} &
\textbf{CLIP-T $\uparrow$} &
\textbf{\#Pass} \\
\hline\hline
Fine-tuning &
0.383 & 0.697 & 0.275 & \textbf{x1} \\
+ CFG &
\textbf{0.529} {\scriptsize\textcolor{green!60!black}{(+38.1\%)}} &
\textbf{0.799} {\scriptsize\textcolor{green!60!black}{(+14.6\%)}} &
\textbf{0.304} {\scriptsize\textcolor{green!60!black}{(+10.5\%)}} &
x2 \\
\cellcolor{gray!10} \texttt{DogFit} &
0.516 {\scriptsize\textcolor{green!60!black}{(+34.7\%)}} &
0.758 {\scriptsize\textcolor{green!60!black}{(+8.8\%)}} &
0.269 {\scriptsize\textcolor{red!60!black}{($-2.2\%$)}} &
\textbf{x1} \\
\hline
\end{tabular}
}
\caption{Comparison on the text-to-image DreamBooth benchmark for fine-tuning Stable Diffusion V1.5. Results are based on the best guidance strength per method ($\omega_{\text{CFG}}=7.5$, $~\omega_{\text{\texttt{DogFit}}}=1.65$).}
\label{tab:results}
\end{table}

\begin{table}[t]
\centering
\scriptsize
\setlength{\tabcolsep}{10pt}
\begin{tabular}{|l|cc|}
\hline
\textbf{Guidance} & \textbf{CLIP-T $\uparrow$} & \textbf{\#Pass} \\
\hline\hline
Real data & \textbf{0.347} & -- \\

Fine-tuning & 0.276 & \textbf{x1} \\
+ CFG        & 0.343 {\scriptsize\textcolor{green!60!black}{(+24.3\%)}} & x2 \\
\rowcolor{black!5} \texttt{DogFit} & 0.319 {\scriptsize\textcolor{green!60!black}{(+15.6\%)}} & \textbf{x1} \\
\hline
\end{tabular}
\caption{Comparison on text-to-image LoRA fine-tuning of Stable Diffusion v1.5 to the Pokémon image–caption dataset. Results are based on the best guidance strength per method ($\omega_{\text{CFG}}=7.5$, $~\omega_{\text{\texttt{DogFit}}}=1.75$).}
\label{tab:lora_results}
\end{table}


\subsection{Scaling to Large Resolutions}
\label{app:scaling-large-res}

Tab.\ref{tab:food512} provides a comparison between \texttt{DogFit} and CFG when fine-tuning DiT-XL/2 pretrained at $512{\times}512$ on Food101~\citep{bossard2014food} at $512{\times}512$. Results indicate that \texttt{DogFit} is fully scalable to larger resolutions, broadening its applicability and demonstrating invariance to image scale. For this experiment, we use a batch size of 12 and use the same hyperparameters as in our main experiments.

\subsection{Robustness to Stronger Domain Shifts}
\label{app:eurosat}

While our target datasets span a range of different distribution gaps, we additionally evaluate transfer on the \textbf{EuroSat} dataset~\citep{helber2019eurosat}, consisting of 27{,}000 satelite images with 10 classes, which shares very few features with the source domain. This setting probes whether \texttt{DogFit}’s source prior still provides a better signal than that of the fine-tuned target under \emph{extreme shift}. We note that as the target domain diverges substantially from the source, the advantages of transfer learning from the source generally diminish relative to training on the target data from scratch. Nevertheless, our results in Tab.\ref{tab:eurosat} show that \texttt{DogFit} still outperforms Fine-tuning and CFG in this case, indicating that leveraging the source prior can still improve generation quality even when transfer benefits are attenuated.  



\subsection{Exponential vs.\ Uniform Scheduling}
Tab.\ref{tab:eurosat} shows that sampling the guidance weight $\omega$ from a SEDD (Fig.~\ref{fig:cdf_}) consistently outperforms uniform schedules across all test-time values of $\omega$. In particular, SEDD yields lower FID while uniform sampling over broader ranges destabilizes learning and collapses diversity, with FID inflating as $\omega$ increases. This empirical trend aligns with our linearity assumption in \textbf{Prop, 1} which hints at the equivalence of \texttt{DogFit} and DoG~\cite{zhong2025domain} under small $\omega$ values~\cite{zheng2023characteristic}. By concentrating the probability mass near $\omega\!\approx\!1$, SEDD keeps optimization within the proposition’s validity region, preserving the expected guidance behavior and allowing the use of larger test-time guidance values during generation.

\subsection{DogFit in Text-to-Image Transfer Learning}
\label{sec:dogfit_t2i_transfer}

To further show the broad applicability of our method, we test \texttt{DogFit} in two different text-to-image transfer settings, Dreambooth subject driven transfer ~\citep{ruiz2023dreambooth}, and LoRA style transfer ~\citep{hu2022lora} to Pokémon. Note that these settings typically require large guidance values of up to $7.5$ for the model to produce visually pleasing images. However, \texttt{DogFit} is more theoretically grounded when exposed to small-guidance values during training, where it locally matches DoG~\citep{zhong2025domain}  (see Prop. 1). While controllable guidance (conditioning on $w$) is a natural extension allowing for large test-time $\omega$ values, we focus here on fixed guidance during training and leave that extension for future work. Despite this conservative setting, \texttt{DogFit} significantly improves unguided fine-tuning with small guidance values ($\omega=[1.65,1.75]$) and approaches the performance of CFG with high guidance value ($\omega=7.5$).


\paragraph{DreamBooth (subject-driven transfer).}
We apply Dreambooth fine-tuning on Stable Diffusion v1.5~\citep{rombach2022high,podell2023sdxl} to novel subjects using a handful of instance images with \texttt{DogFit}. In this regime, the fine-tuning prompts are typically limited to templates such as \emph{“a [v] photo of \{\textit{subject}\}”}, while at inference we expect the model to respond to diverse prompts from the source domain. This mismatch makes faithful distillation of text-conditioning challenging. Nevertheless, Fig.~\ref{fig:dreambooth_qualitative} and Tab.~\ref{tab:results} show that \texttt{DogFit} learns controllable alignment with the prompt and significantly outperforms unguided fine-tuning while approaching CFG on DINO, CLIP-I, and CLIP-T, the standard metrics for subject-driven generation~\citep{ruiz2023dreambooth}.

\paragraph{LoRA fine-tuning on Pokémon (style transfer).}
We adapt Stable Diffusion~v1.5 to the Pokémon style dataset\footnote{https://huggingface.co/datasets/diffusers/pokemon-gpt4-captions} with 833 image–caption pairs.
We train a LoRA adapter (rank 4) for 2{,}000 iterations at a learning rate of $10^{-4}$. We evaluate with CLIP-T over the first 100 prompts (one image per prompt). As summarized in Tab.\ref{tab:lora_results}, \texttt{DogFit} significantly improves prompt alignment over unguided fine-tuning while using a small guidance value ($\omega{=}1.75$) and a single sampling pass, and it closely approaches CFG which requires two passes and a large guidance value ($\omega{=}7.5$).

\end{document}